\pdfoutput=1
\documentclass[reprint,amsmath,amssymb,aps,showkeys]{revtex4-2}

\usepackage{graphicx}
\usepackage{dcolumn}
\usepackage{bm}
\usepackage{xcolor}
\usepackage{hyperref}

\usepackage{tikz}
\usepackage[T1]{fontenc}
\usetikzlibrary{positioning, shapes.geometric, intersections}

\usepackage{booktabs,siunitx,array,threeparttable}

\newcommand{\bs}[1]{\boldsymbol{#1}}
\newcommand{\In}{\mathcal{I}n}
\newcommand{\Av}{\mathcal{A}v}

\newcommand{\newtext}[2]{#2}
\newcommand{\rev}[2]{#2}

\definecolor{blush}{rgb}{0.87, 0.36, 0.51}
\hypersetup{
    colorlinks,
    linkcolor={blush},
    citecolor={blue!70!black},
    urlcolor={blue!70!black}
}

\begin{document}

\preprint{APS/123-QED}

\title{Dimensionless Numbers Reveal Distinct Regimes in the Structure and Dynamics of Pedestrian Crowds}

\author{Jakob Cordes}
\email{j.cordes@fz-juelich.de}
\altaffiliation[Also at ]{Institute of Advanced Simulation, Forschungszentrum J\"ulich GmbH, J\"ulich, Germany}

\author{Andreas Schadschneider}
\email{as@thp.uni-koeln.de}
\affiliation{ Institut f\"ur Theoretische Physik, Universit\"at zu K\"oln, Germany}

\author{Alexandre Nicolas}
\email{alexandre.nicolas@cnrs.fr}
\affiliation{Institut Lumi\`ere Mati\`ere, Universit\'e Claude Bernard Lyon 1 and CNRS, France}

\date{\today}

\begin{abstract}
In Fluid Mechanics, dimensionless numbers like the Reynolds number help classify flows. We argue that such a classification is also relevant for crowd flows by putting forward the dimensionless Intrusion and Avoidance numbers\newtext{}{, which quantify the intrusions into the pedestrians' personal spaces and the imminency of the collisions that they face, respectively.} 
Using an extensive dataset, we show that these numbers delineate regimes where distinct variables characterize the crowd's arrangement, namely, Euclidean distances at low Avoidance number and times-to-collision at low Intrusion number.
\newtext{}{On the basis of these findings, a perturbative expansion of the individual pedestrian dynamics is carried out around the non-interacting state, in quite general terms. Simulations confirm that this expansion performs well in its expected regime of applicability.}
\end{abstract}


\maketitle


\section*{Introduction}
Crowds often look like an ocean made of hundreds or thousands of heads, ruffled by ripples and waves \cite{bain_DynamicResponseHydrodynamics_2019}, moving in synchrony or not; this impression struck poets \cite{hugo1872annee} long before it inspired scientists \cite{henderson1974fluid,hughes2003flow}. Yet, even as of now, pedestrian dynamics as a discipline does not stand on the same footing as fluid mechanics. 
In the latter field, the classical motion of particles at the microscale is governed by an exact equation, Newton's law, whose homogenization yields the universal Navier-Stokes equation. In practice, modelers resort to a plethora of approximate schemes (e.g., Stokesian dynamics, lattice Boltzmann methods, Euler equations for inviscid flows) but the choice among these is guided, and theoretically bolstered, by the calculation of dimensionless numbers, such as the Reynolds number and the Mach number.

On the other hand, a zoo of models for pedestrian dynamics co-exist (see e.g. \cite{Martinez-Gil2017,SchadschneiderCSTZ18,ChraibiTSS18,maury_CrowdsEquationsIntroduction_2018}) and the realm of applicability of each is ill-defined. The crowd's density is generally used to delineate different regimes, for instance the
levels of service defined by Fruin for crowds  \cite{Fruin,schadschneider2010stochastic}.
Each level is marked by a dominant behavior: (un)avoidable contact, necessity to change gait, possibility to turn around, etc. and it has been argued that as the density changes crowd dynamics should be controlled by distinct laws \cite{best_DenseSenseInteractiveCrowd_2014}. However, the watersheds between the regimes are arbitrary.
Even from a practical standpoint, for safety assessments, crowds at similar densities may present contrasted characters and risk profiles.
Consider the difference between a densely packed, but static audience in a concert hall and people vying for escape in an emergency evacuation \cite{FelicianiBook}.
Recently, yearning for a better classification of these scenarios, it was proposed to gauge congestion on the basis of a dimensionless number related to the vorticity of the velocity field, instead of the density \cite{zanlungo_PureNumberAssess_2023}.
This quantity is practically relevant, notably for safety issues, but gives no insight into the determinants of pedestrian dynamics at the microscale.

In this paper, we argue that in common scenarios pedestrian dynamics are dominated by two variables, rendering the ideas of preservation of personal space (proxemics) and anticipation of collisions. Their averages over the crowd define dimensionless parameters that delineate regimes of crowd flows which display distinctive features. In the spirit of the Reynolds number in fluid mechanics, these dimensionless numbers help gauge the range of validity of pedestrian models and prompt specific perturbative expansions of the equations of motion in each regime.


\section*{Results}
Psychological studies on proxemics indicate that people pay attention to their personal space, defined as \emph{"the area individuals maintain around themselves into which others cannot intrude without arousing discomfort"} \cite{hayduk_PersonalSpaceEvaluative_1978}, more than to global density \cite{evans2007crowding}. We are therefore led to define an intrusion variable $\In_i$ centered on each agent $i$, which vanishes for isolated pedestrians and, ideally, diverges at physical contact, so as to underscore the transition from no-contact dynamics to contacts and pushes. 
Here, to keep at bay ambiguities in the definition of a local density, we make use of the center-to-center distances $r_{ij}$ to define  $\In_i$, viz.,
\begin{equation}
    \In_i = \sum_{j\in \mathcal{N}_i} \underset{ \In_{ij} }{\underbrace{\left(  \frac{r_\text{soc}-\ell_\text{min}}{r_{ij}-\ell_\text{min}}  \right) ^{k_I}}  }.
            \label{Eq:In}
\end{equation}
Taking $k_I=2$, this represents the sum of \emph{areal} encroachments of other agents $j$ on $i$'s personal space.

For simplicity, we overlook anisotropic effects and assume uniform circular shapes for the pedestrian bodies and personal spaces, of diameter $\ell_\text{min}=0.2\,\mathrm{m}$ and radius $r_\text{soc} = 0.8$~$m$, respectively.
The sum runs over the set $\mathcal{N}_i$ of all close neighbors $j$ of $i$, here defined by $r_{ij} \leqslant 3 \,r_\text{soc}$. 
That the intrusions of diverse neighbors should be added up makes sense for physical contacts (superposition of mechanical forces), but also for proxemic behavior  \cite{lian2018analysis, knowles_GroupSizeExtension_1976}.

While this variable gives a sense of the level of crowding, it neither provides a full reflection of psychological experience (feeling of congestion) in the midst of the crowd \cite{jia2022revisiting, zanlungo_PureNumberAssess_2023}, nor fully controls the agent's dynamics: when two people $i$ and $j$ run towards each other, they will not behave as though they were isolated, even though they may still be separated by several meters, hence, $\In_{ij} \to 0$. This anticipatory behavior is well captured by an anticipated time-to-collision (TTC) $\tau_{ij}$, defined as the delay until the first collision if both $i$ and $j$ keep their current velocities ($\tau_{ij}=\infty$ if no collision is expected).
Humans are indeed capable of identifying the most imminent collision between multiple objects and estimating TTCs \cite{delucia_JudgmentsRelativeTimetocontact_1997}, notably via purely optical quantities, namely, the optical angle divided by its derivative \cite{lee_TheoryVisualControl_1976}.
Experiments showed that the TTC is instrumental in humans' decisions as to `when' to avoid an approaching pedestrian \cite{pfaff_AvoidanceBehavioursYoung_2018}. 
Accordingly, the TTC can be used to define a non-dimensional avoidance variable $\Av_i$
quantifying the risk of an imminent collision,
\begin{equation}
    \Av_i = \sum_{j \in \mathcal{N}_i^{\prime}} \underset{\Av_{ij}}{\underbrace{\Big(\frac{\tau_0}{\tau_{ij}}\Big)^{k_A}}}.
     \label{Eq:av}
\end{equation}
Here,  $\tau _0=3\,\mathrm{s}$ is a timescale  above which collisions are hardly dreaded. The exponent $k_A$ is simply set to 1 (see Appendix~B \newtext{}{in the supplemental material} for variations), 
and, in contrast with Eq.~\ref{Eq:In},  the set of neighbors $\mathcal{N}_i^{\prime}$ is restricted to the agent with the shortest $\tau_{ij}$, i.e., the most imminent risk. Indeed, for collision avoidance, it has been ascertained that participants immersed in a virtual crowd tend to fixate a particular agent with a high risk of collision just before performing an avoidance maneuver around this person \cite{meerhoff2018guided}.


\begin{figure*}
    \centering
    \includegraphics[width = \linewidth]{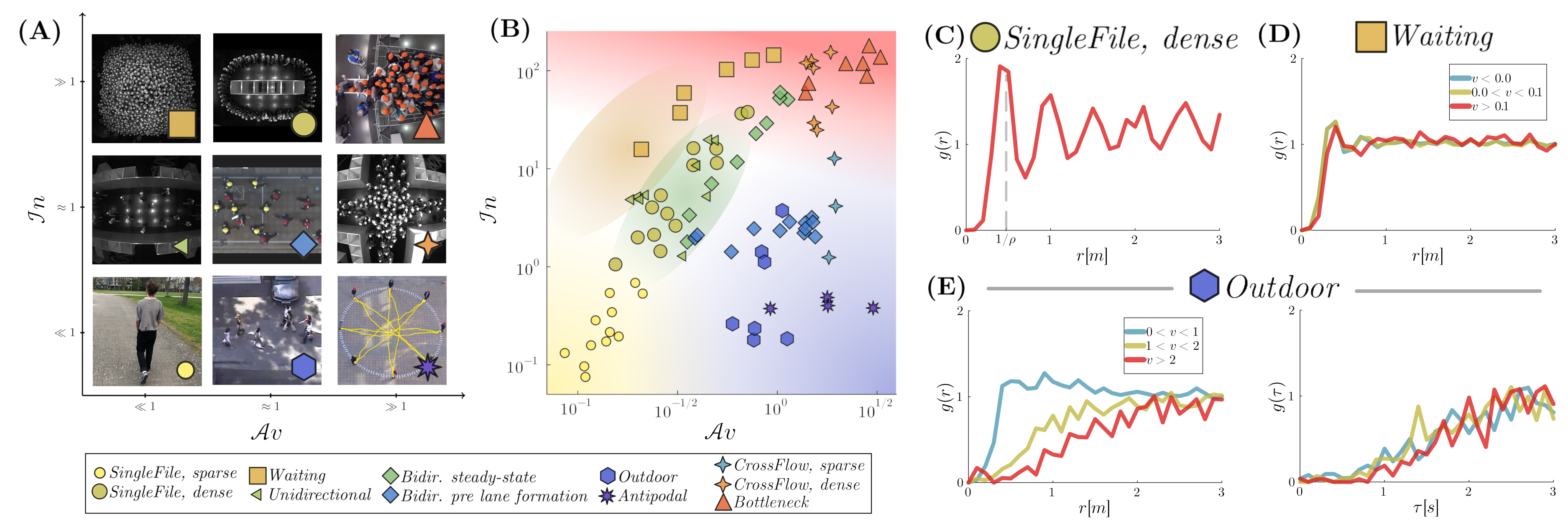}
    \caption{Delineation of crowd flow regimes with the dimensionless numbers $\In$ and $\Av$. 
    (\textbf{A}) Sketch of the expected diagram, illustrated with snapshots from the collated datasets (described in Appendix~A \newtext{}{in the supplemental material}). 
    (\textbf{B}) Empirical diagram obtained from various pedestrian datasets.
    Each datapoint corresponds to one experimental run or observational sequence.
    The experimental data from \emph{SingleFile} and \emph{CrossFlow} were split into a \emph{sparse} and a \emph{dense} dataset. 
    Coloured gradients are visual guides, to indicate different regimes.
    (\textbf{C}) Pair-distribution function (pdf) for one run of the \emph{SingleFile, dense} dataset.
    \textbf{(D)} pdf for the \emph{Waiting} dataset and (\textbf{E}) for the \emph{Outdoor} dataset.
    The curves are binned according to the rate of approach $v = - dr/dt$ (given in $m/s$).
    As for (\textbf{D}) the pdf is well parameterised by $r$ and poorly by $\tau$ (see Fig.~$3$ in Appendix~C  \newtext{}{in the supplemental material}) and in \textbf{(E)} vice versa.
    }
\label{fig:combined_figure}
\end{figure*}


With these two variables in hand, one can hope for a finer delineation of pedestrian streams than with the traditional density-based levels of service. To this end, the foregoing agent-centered variables are averaged over the $N(t)$ agents observed in the crowd at time $t$, and then over time. This average defines the dimensionless Avoidance number $\Av$ and Intrusion number $\In$. As $\Av$ should quantify the urgency of expected collisions we only consider data points with a finite TTC in the average. Especially in the sparse datasets, this allows to focus on the parts where interactions occur.

Figure~\ref{fig:combined_figure}A illustrates the regimes of crowd flow that one would intuitively expect to find in a diagram parametrized by $\Av$ and $\In$, using exemplary cases.
The bottom left corner, $\In, \Av \ll 1$, corresponds to very sparse crowds with hardly any interactions.
As one moves up the $\In$-axis, the setting gets more crowded, and pedestrians are eager to maintain a certain social distance with respect to others, as in a \emph{unidirectional} flow. 
When $\In \gg 1$, personal space can no longer be preserved and physical contact may eventually be unavoidable, as in a tightly packed static crowd (\emph{Waiting} scenario).
A very different way to depart from the non-interacting case is to consider people walking or running towards each other. This is well approximated by the beginning of an \emph{Antipodal} experiment, in which participants initially positioned all along the circumference of a circle (with $\In \ll 1$) are asked to reach the antipodal position. This induces conflicting moves, with risks of collision in the center of the circle, hence $\Av \gg 1$.
Finally, competitive evacuations though a \emph{bottleneck} exemplify the regime of large $\In$ and $\Av$, which features contacts, pushes, as well as conflicting moves.

These are of course idealized expectations. To test them, we have collated an extensive dataset of pedestrian trajectories, including controlled experiments (single-file motion \cite{seyfried_PhaseCoexistenceCongested_2010}, bottleneck flows \cite{adrian_CrowdsFrontBottlenecks_2020}, corridor flows \cite{cao_FundamentalDiagramsMultidirectional_2017, feliciani_UniversalFunctionCapacity_2018, murakami_MutualAnticipationCan_2021}, antipodal scenarios \cite{xiao_InvestigationPedestrianDynamics_2019}) and empirical observations in outdoor settings \cite{pellegrini_YouLlNever_2009, lerner_CrowdsExample_2007a}; further details about these scenarios and the way we have smoothed out head sways from the trajectories can be found in Appendix~A \newtext{}{in the supplemental material}. For each scenario and each realization, we have computed $\Av(t)$ and $\In(t)$ every $0.5\,$s, and averaged over the whole quasi-stationary state (unless otherwise stated).

Figure~\ref{fig:combined_figure}B shows that the idealized diagram worked out intuitively (Fig.~\ref{fig:combined_figure}A) is largely corroborated by the empirical datasets.
Indeed, single files of amply spaced pedestrians are found in the bottom left corner, at small $\In$ and $\Av$, whereas the top of the diagram, at large $\In$, is occupied by situations in which physical contacts are almost inevitable. More interestingly, unidirectional flows and cross-flows may have similar $\In$ numbers, but they are distinguished by $\Av$, which takes larger values for cross-flows, prone to more conflicts. In the same vein, antipodal maneuvers have intrusion numbers comparable to those of some typical outdoor scenarios, but larger avoidance numbers.
The spread of points for a given type of scenario is expected and sensible, as a given geometry can give rise to flows of different natures, depending on the inflow and density, notably. Conversely, scenarios bearing different names may be similar on the whole and thus have comparable dimensionless numbers; for instance, a bidirectional flow with thick lanes is mostly made of unidirectional flows, apart from the few interactions between the lanes.
Note that the $\In$ and $\Av$ axes have been plotted orthogonally, whereas skewed axes should in principle be used if the variables exhibit some correlations; this does not alter the topology of the diagram, however. Nor do variations of the (somewhat arbitrary) precise definitions of $\In$ and $\Av$, see Appendix B \newtext{}{in the supplemental material}.
\newtext{}{In particular, $\In$ is related to, but yet different from, the density, in that it is associated with individuals and captures both psychological and bio-mechanical aspects, as further discussed in Appendix~B.}

In practice, the visual delineation of regimes on the diagram of Fig.~\ref{fig:combined_figure}B appears sensible. 
But its physical relevance will only transpire if the delineated regimes exhibit constitutive differences. Remarkably, we find a major difference in the arrangement of the crowd, not in terms of static symmetry of the structure (which distinguishes, say, a liquid from a crystal), but in the nature of this self-organized `structure', i.e., more pragmatically, in the variables that characterize it. Drawing inspiration from condensed matter physics and following \cite{karamouzas_UniversalPowerLaw_2014}, we use as structural probe the pair-distribution function (pdf) $g(x)=P(x)/P_{\text{NI}}(x)$ between pedestrians, which quantifies the probability that two interacting pedestrians are found a given distance $x$ apart, renormalized by the probability $P_{\text{NI}}$ of measuring this distance for pedestrians that do not interact. This probability can be approximated by randomizing the time or space information (cf. Appendix~C \newtext{}{in the supplemental material}).

Starting from the origin ($\In, \Av = 0$) and moving up along the $\In$-axis while keeping $\Av \ll 1$, the crowd gets structured in real space, as evidenced by its radial pdf $g(r)$, where $r$ is the Euclidean spacing between people.
This is conspicuous for one-dimensional configurations; indeed, the pdf of dense single files  (Fig.~\ref{fig:combined_figure}C) develops a series of gradually decaying oscillations, with peaks positioned at multiples of the mean spacing, resembling the pdf of a liquid or a dense suspension of active colloids \cite{klongvessa2020study}.
But structural features are also visible in two-dimensional settings, notably the dense static waiting crowd (Fig.~\ref{fig:combined_figure}D). 
Its pdf displays a strong dip at short distances, below $0.3\sim 0.4\,\mathrm{m}$, reflecting strong short-range repulsion, due to hard-core impenetrability and the reluctance for intrusion into the intimate space; the dip is followed by a peak at the nearest-neighbor distance. These features in real space are insensitive to dynamic variables such as the rate of approach $v$ (i.e., the rate at which the distance between two pedestrians declines): the radial pdf exhibit the very same trend (Fig.~\ref{fig:combined_figure}D), quite independently of $v$.

The situation is widely different if one departs from the non-interacting regime by turning up $\Av$, i.e., considering very sparse crowds ($\In \ll 1$) with more and more conflicting moves, as in the antipodal scenario or sparse outdoor crowds. This is the regime analyzed in \cite{karamouzas_UniversalPowerLaw_2014}.
Strikingly, the radial pdfs do not collapse onto a single curve in this case; binned by rates of approach $v$, their pdfs display different shapes (left of Fig.~\ref{fig:combined_figure}E).
In particular, the faster pedestrians approach each other, the larger is the Euclidean spacing at which they begin to interact.

Instead, if the TTC $\tau$ is substituted for $r$ as the argument of the pdf, then a master curve is recovered, as shown in  Fig.~\ref{fig:combined_figure}E (right) for the \emph{Outdoor} dataset.
In particular, the pdf gets more and more strongly depleted as $\tau$ becomes shorter, signalling the risk of an imminent collision.
Thus, crowds in this regime also have some structure, but this is mostly hidden in real space and only becomes apparent in TTC space.
This major finding of \cite{karamouzas_UniversalPowerLaw_2014} is here contextualized by ascribing it to a particular regime of crowd flow: it does not hold for e.g. the waiting room (finite $\In$, small $\Av$) (Fig.~$3$ in Appendix~C \newtext{}{in the supplemental material}).



\section*{Discussion}

To what extent can these observations be rationalized theoretically?
Formally, the dynamics of a pedestrian $i$ (or any other entity) is a function of their perceived surroundings, more precisely, the set $\mathcal{R}(t)=\left(\bs{r}_1(t),\dots,\bs{r}_N(t)\right)$  of all positions of the $N$ agents (and, if need be, body orientations) observed so far,  the agents' shapes $\mathcal{S}$, and some variable $\xi_i$ gathering all unobserved features, which (in the worst case) may vary from realization to realization. Without loss of generality, 
\newtext{this functional dependence for each agent can be expressed as a minimization over a suitably defined cost function $\mathcal{C}_i$ \cite{VanToll2020,HoogendoornB03,guy_LeasteffortTrajectoriesLead_2012},}
{it is possible to recast this functional dependence as a minimization, by \emph{designing} a suitable mathematical function $\mathcal{C}_i$ (hereafter called cost function to follow the common terminology \cite{VanToll2020,HoogendoornB03,guy_LeasteffortTrajectoriesLead_2012}, but with no implication on its nature or properties), viz., }
\begin{equation}
    \bs{v}_i^\star= \underset{ \bs{v} \in \mathbb{R}^2 }{\arg \min}\,\mathcal{C}_i\Big[\bs{v}, \{\mathcal{R}(t^{\prime}), t^{\prime}\leqslant t\}, \mathcal{S}, \xi_i\Big]
    \label{eq:dynamics1}
\end{equation}
where $\bs{v}_i^\star$ denotes the decision of agent $i$ which serves as an input to a mechanical layer which yields the actual velocity $\bs{v}_i(t)$. Unfortunately, neither the cost function $\mathcal{C}_i$ nor the hidden variables $\xi_i$ are known.
Nevertheless, in the physical sciences, generic perturbative expansions
often afford ground for the study of systems near their critical states, leveraging symmetries to compensate blurred microscopic insights. Here, one cannot rely on conventional symmetry considerations,
but the empirical classification of crowd regimes performed above has confirmed the prominent role of the Intrusion and Avoidance number.
This justifies the assumption that the agents' responses are mostly controlled by $\In_i$ and $\Av_i$, so that the second part of Eq.~\ref{eq:dynamics1} reduces to 
\begin{equation}
    \bs{v}_i^\star = \underset{ \bs{v} \in \mathbb{R}^2 }{\text{arg min}}\,\mathcal{C}_i\left[\bs{v}, \In_i\left(\bs{r}_i(t) + \bs{v} \delta t\right), \Av_i (\bs{v}) \right],
    \label{eq:dynamics2}
\end{equation}
where $\Av_i$ (i.e., the TTC) is evaluated with the test velocity $\bs{v}$ and $\In_i$ at the associated position $\bs{r}_i(t) + \bs{v} \delta t$ where $\delta t$ is a time step. 
In the non-interacting scenario ($\In_i, \Av_i =0)$, the agent freely pursues her goal at velocity $\bs{v}_{\text{des}, i}$, hence, to leading order,
\begin{equation}
    \mathcal{C}_i(  \bs{v} )  \approx (\bs{v}_{\text{des}, i} - \bs{v} )^2.
\end{equation}
Expanding Eq.~\ref{eq:dynamics2} around this reference situation, as detailed in Appendix~D \newtext{}{in the supplemental material}, yields
the $\Av*\In$-model,
\begin{widetext}
\begin{equation}
    \mathcal{C}_i\left[\bs{v}, \In_i\left(\bs{r}_i(t) + \bs{v} \delta t\right), \Av_i (\bs{v}) \right]  \approx  \left[ \bs{v}_{\text{des}, i}  - \bs{v}  + \beta \nabla \In _i \left(\bs{r}_i(t)\right) \right]^2 + \alpha\, \Av_i (\bs{v}) 
    \label{eq:AV_IN_Model}
\end{equation}
\end{widetext}
with $\alpha, \beta \geq 0$. We will refer to the case $\alpha = 0$ as the $\In$-model and $\beta = 0$ as the $\Av$-model.
\newtext{}{One should bear in mind that these models were derived as generic asymptotic expansions of Eq.~\ref{eq:dynamics2}, rather than designed in an \emph{ad hoc} way;}
\newtext{}{their purpose is not so much to be realistic for the widest possible range of scenarios, as it is to shed light on crowd dynamics in limiting cases. In Eq.~(\ref{eq:AV_IN_Model}) }
we have neglected all mechanical interactions between the agents and the actual velocity relaxes towards the optimum $\bs{v}_i^\star$ over a time-scale $\tau _R$.

Let us test this perturbative expansion in the corresponding (asymptotic) regimes.
First, we simulate the \emph{Waiting} scenario:
In the $\In$-model, the agents make use of the available space to keep social distances to the others, which results in comparable averaged $\In$ numbers ($\In=16$ for the experiments vs.\ \newtext{}{$\In=14$} with the $\In\text{-model}$).
By contrast, the $\Av$-model fails to capture these features: the system remains frozen in its initial state as no collision is expected.
The central role of $\In$ is also readily understood in the case of a waiting line, where people halt to preserve each other's personal space.
As a consequence, in a macroscopic model of the crowd, the local flow will depend solely on the density field, echoing the finding of a density-based hydrodynamic response of the crowd at the start of a marathon   \cite{bain_DynamicResponseHydrodynamics_2019}.
In the opposite regime, the basic features of the sparse \emph{CrossFlow}, notably successful collision avoidance, are well replicated by the $\Av$-model, contrary to the $\In$-model in which the agents bump into each other. They are unable to maintain reasonable spacings (in TTC or in real space) with respect to each other, as also testified by the values of the dimensionless numbers (we find $\Av=$ 1.8 / 1.1 / 6.8 and $\In=$ 1.2 / 2.9 / 5.3 in the experiments / $\Av$-model /  $\In$-model, respectively).


\begin{figure}
\centering
    \includegraphics[width= 0.8 \linewidth]{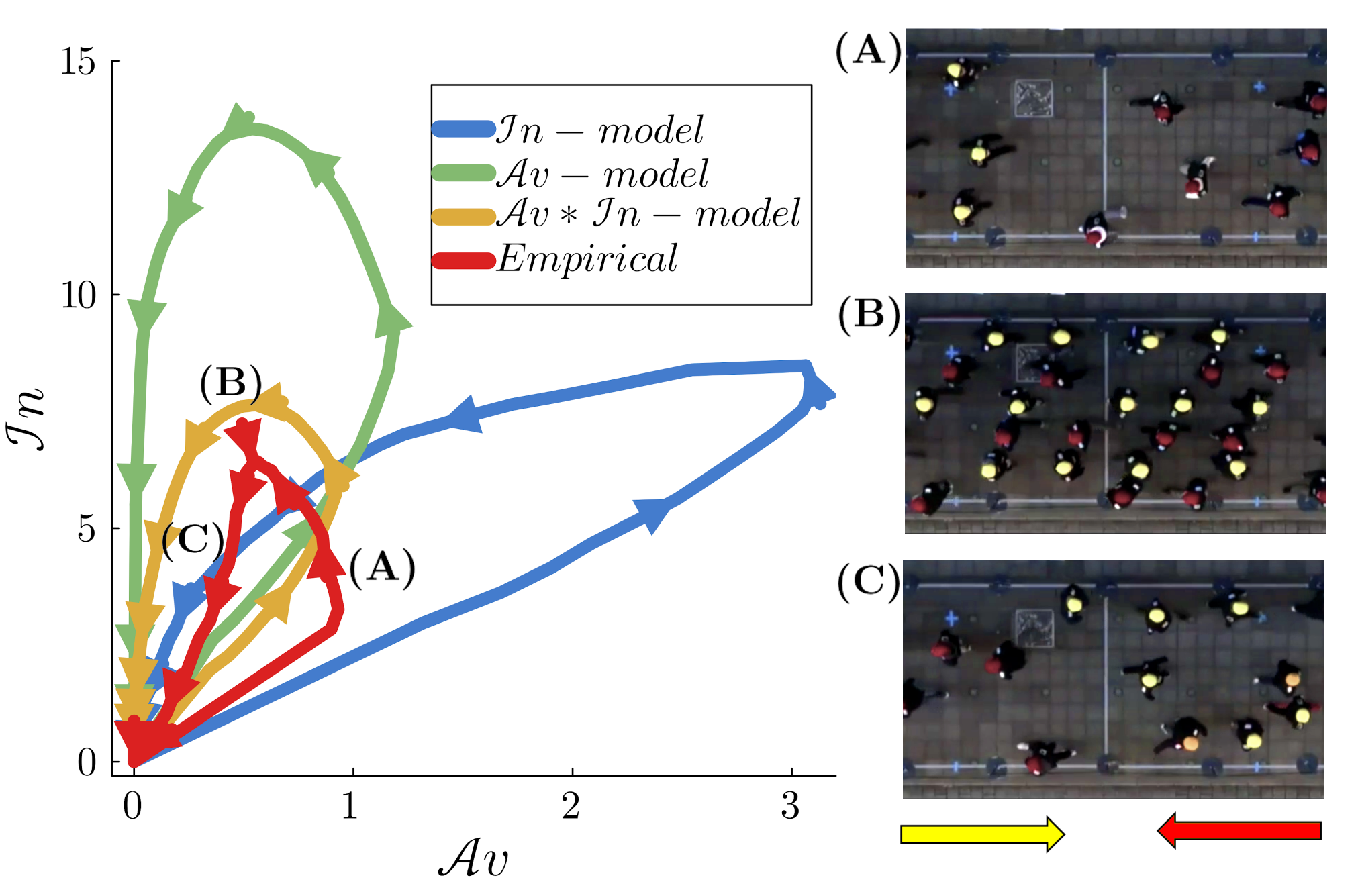}
\caption{Phase space trajectory during and after the formation of lanes using empirical results 
\cite{murakami_MutualAnticipationCan_2021} and simulations of the proposed models. The arrows in the $\In-\Av$ plot are spaced by $2.5\mathrm{s}$. The temporal values of $\Av$ and $\In$ are averaged over multiple realizations of the experiment as well as the simulations, with random initial conditions.}
\label{fig:phase_space_trajectory}
\end{figure}


The deficiency of models premised solely on $\In$ or $\Av$ is even more manifest in scenarios which are not confined to the vicinity of the axes of the $(\Av,\In)$ plane.
For example, let us pay attention to the temporal evolution of a bidirectional flow, using as input the experimental data of \cite{murakami_MutualAnticipationCan_2021} and averaging over multiple similar realizations.
The process of lane formation and then disappearance of the lanes after the two groups have passed each other entails a loop in the phase space, as represented in Fig.~\ref{fig:phase_space_trajectory}. 
Shortly after pedestrians enter the measurement area, in \textbf{A}, the limited space for each crowd leads to moderate values of $\In$, but $\Av$ gets relatively high as the groups are walking towards each other, until they form lanes in \textbf{B}, thus lowering $\Av$, while $\In$ is large because space is limited; finally, in \textbf{C}, the crowds have passed each other (low $\Av$) and the pedestrians make use of the available space by dissolving the lanes (moderate $\In$), marking a return to the origin.
Even though all models reproduce the formation of lanes, only the $\Av*\In$-model produces a loop comparable to the empirical one.
While the $\In$-model is unable to keep in check the growth of $\Av$ prior to lane formation, the $\Av$-model fails to ensure sufficient space between people when lanes have formed, leading to very high $\In$ values.
The dynamics of all scenarios are shown in the supplemental video.

While the focus was here put on the asymptotic $\In$ and $\Av$-models, the discussion has bearing on the broader category of agent-based models: their equations of motion often hinge on variants of either the $\In_{i}$ variable \cite{helbing_SimulatingDynamicalFeatures_2000, tordeux_CollisionFreeSpeedModel_2016} or the $\Av_{i}$ variable \cite{vandenberg_ReciprocalVelocityObstacles_2008,vandenberg_ReciprocalNBodyCollision_2011,karamouzas_UniversalPowerLaw_2014}
(for instance, the \emph{interactions} in the models of \cite{vandenberg_ReciprocalVelocityObstacles_2008} and \cite{karamouzas_UniversalPowerLaw_2014} boil down to those of the $\Av$-model with $k_A=1$ or $2$ in Eq.~\ref{Eq:av}, respectively),
thereby limiting their range of applicability to the associated regime; a detailed inspection of this broader model category is deferred to a future publication.

Finally, in all regimes discussed so far, contacts between pedestrians were  at most scarce.
The situation is different in the high-$\In$ region, which is highlighted in red in Fig.~\ref{fig:combined_figure}B and notably includes competitive bottleneck flows; in that case, more realistic (e.g., elliptic) shapes and mechanical contacts should be considered.


\section*{Conclusion}

In summary, we have shown that the desire to preserve one's personal space from intrusions and the anticipation of collisions, quantified by the dimensionless numbers $\In$ and $\Av$, delineate different regimes at the crowd's scale. These are marked by specific dynamics and `structural' arrangements. The importance of taking into account these factors to model the dynamics of individual agents depends on the regime under study.

At present, only collisions between the hard-cores have been taken into account, in the absence of which ($\Av_i  = 0$), agents are deemed isolated and have thus been left aside in the averaged $\Av$. 
In reality, the 'softer' collisions, i.e. the  anticipated  intrusions into the private or intimate space, are also avoided.
A more sophisticated definition of $\Av$ should be able to capture these.

Beyond $\In$ and $\Av$, other dimensionless numbers can, and certainly should, be introduced to describe specific features of crowd dynamics such as  an analogue of the Mach number for the propagation of waves in crowds or some variant of the Péclet number (diffusion over advection) to account for the variability in the outcome of nominally similar experiments, due to the hidden variables $\xi_i$ in
Eq.~\ref{eq:dynamics1}.
Interestingly, such a series of dimensionless numbers would mark successive departures from the conservation laws and invariance principles traditionally encountered in physical systems: while in the $\In$-regime agents do not differ from particles subjected to distance-based interactions, $\Av$ introduces a velocity-based component to the interactions and a marginal violation of the reciprocity of forces.
Better capturing the asymmetry of perception between pedestrians would make the violation of reciprocity more acute, with all its implications in active systems \cite{fruchart2021non}.
Eventually, the violation of Galilean invariance in crowds would be mirrored by paying attention not only to TTC, but also to absolute time gaps, should the neighbors suddenly come to a halt.
By gradually relaxing the symmetries applicable in physical systems, the way is thus paved for a general theoretical study of the statistical physics of pedestrian assemblies. 
\newtext{}{In particular, it may be a good strategy to first focus on regimes where many of the aforementioned numbers are zero in order to derive a macroscopic flow theory starting from the Boltzmann equation \cite{bertin2006boltzmann,arnault2022chapman} (along the lines propounded by Chapman and Enskog for fluids), before facing the specific challenges raised by nonzero dimensionless numbers (e.g., the nonlocality inherent in the $\Av \neq 0$ regime). }


\section*{Materials and Methods}

\paragraph*{Empirical Datasets.}
Each data set contains trajectories expressed in real-world coordinates as a function of time.
For the \emph{Waiting} scenario we extracted trajectories by ourselves using the semi-automatic tracking mode of the \emph{PeTrack} software \cite{boltes_AutomaticExtractionPedestrian_2010}, from existing videos of a controlled experiment conducted in Germany in 2013 in the frame the \emph{BasiGo} project.

The \emph{Single-File} and the \emph{Cross} dataset were split into a \emph{dense} part and a \emph{sparse} part (comprising the $6$ and $3$ runs with the lowest global densities). 
The initial transients were discarded in most scenarios, but not for the bidirectional flows, where they were used to probe the situation before lane formation. 

\paragraph*{Data processing.}
Pedestrian trajectories typically feature oscillations due to head sways and empirical noise, which both affect the calculation of $\In$ and above all $\Av$. To smooth out the head sways, a $4$-th order Butterworth filter with critical frequency $0.5$~Hz was applied to the trajectories. Velocities were then computed as the distance covered in approximately $1\,\mathrm{s}$. Then, the TTC was computed by assuming that each pedestrian is a disk of diameter $\ell = 0.2\,\mathrm{m}$. This may generate apparent overlaps; to mitigate these artifacts, we set an upper bound $\In _{ij}^{\text{max}} = 400$ and $\Av _{ij}^{\text{max}} = 60$ on all variables for the computation of the $\In$ and $\Av$ numbers.

Some scenarios, particularly the \emph{Outdoor} one, involve a large number of pedestrians that actually walk in isolation. To focus on actual interactions in the assessment of $\Av$,  pedestrians with $\Av_i=0$ were excluded.

\paragraph*{Pair distribution functions (pdf).}
For some variable $x$ ($x=r$ or $x=\tau$ in the following), the pdf is given by the probability that two pedestrians are separated by $x$ normalized by the probability $P_{\text{NI}}(x)$ that two non-interacting pedestrians are separated by $x$, viz., $g(x) = P(x)/P_{\text{NI}}(x)$. This normalization is aimed at correcting the lack of translational invariance in crowd observations.
In practice, $P(x)$ is directly estimated from the frequency of occurrence of `separation' $x$ in the dataset. $P_{\text{NI}}(x)$ is unknown in principle, but can be estimated by randomizing either the spatial or the temporal information \cite{karamouzas_UniversalPowerLaw_2014}, i.e., reshuffling pedestrians and frames.
Finally, the pdf is obtained by binning the data into bins of size $0.1\,\mathrm{m}$ (for $x=r$) or $0.1\,\mathrm{s}$ (for $x=\tau$).

\paragraph*{Asymptotic models.}
The $\Av$, $\In$, and $\Av*\In$-models defined in the main text were simulated in \emph{Julia}, by taking uniform and constant model parameters: $\alpha = 1.5\,\mathrm{m^2/s^2}$, $\beta = 0.02\,\mathrm{/s}$, $v_{\text{des}} = 1.4\,\mathrm{m/s}$, and  $\tau _{R} = 0.1\,\mathrm{s}$. Speeds were capped at $v_{\text{max}} = 1.7\,\mathrm{m/s}$. Agents are modeled as hard disks of radius $\ell_{\text{min}}=0.2\,\mathrm{m}$. Nevertheless, to account for the fact that people shun collisions not only between their hard cores, but also between their private spaces, the radius
was increased to $\ell_{\text{soc}} = 0.4\,\mathrm{m}$ for the computation of $\Av_i$.
Also note that a small scalar $\varepsilon>0$ is subtracted from $\In_{ij}$ in Eq.~\ref{Eq:In} to make $\In_i$ continuous across the cut-off distance.


\begin{acknowledgments}
We are grateful to Maik Boltes for giving us access to the Waiting Room experimental data and helping us in the process of extracting the trajectories. 
We also thank Antoine Tordeux and Mohcine Chraibi for insightful discussions.

The authors acknowledge financial support from the German Research Foundation (Deutsche Forschungsgemeinschaft DFG, grant number 446168800) and the French National Research Agency (Agence Nationale de la Recherche, grant number ANR-20-CE92-0033), in the frame of
the French-German research project MADRAS.
\end{acknowledgments}

\section*{Data Availability Statement}
\newtext{}{The data supporting the findings of this study are available in the supplementary materials for the \emph{Waiting} scenario.
The data from the \emph{Antipodal} scenario have been made available upon request from the authors of \cite{xiao_InvestigationPedestrianDynamics_2019}.
The remaining data are openly available in the repositories listed in Appendix~A in the supplemental material.}

\newpage

\renewcommand{\theequation}{S\arabic{equation}}
\renewcommand\thesubsection{\arabic{subsection}}

\onecolumngrid

\section*{Appendix A: Description and curation of the datasets}
\label{appendix_A}

A large collection of mostly openly available data on pedestrian dynamics has been used in the main text.
The collated data and the methods employed to pre-process and extract average dimensionless numbers $\In$ and $\Av$ from them are detailed in this section.

\subsection{Summary of the datasets}


\begingroup
\setlength{\tabcolsep}{6pt}
\renewcommand{\arraystretch}{1.0}

\begin{table}[!h]
\centering
\begin{threeparttable}
\begin{tabular}{l l l l l l l}
\toprule
{Name} & {Type} & {Scenario} & {Varied Parameter} & {Details} & {Cit.} & {Data} \\

\midrule
\emph{Waiting} & Exp.  & Static & Density &   J\"ulich, Ger, 2013 & - & Supp. \\ 
\emph{Single-File} & Exp. & Single-File & Density &   J\"ulich, Ger, 2006 & \cite{seyfried_PhaseCoexistenceCongested_2010} &  \cite{_DatabaseJulich_}\\ 
\emph{Unidirectional I} & Exp. & Uni-dir.  & Density &  J\"ulich, Ger, 2013 & \cite{cao_FundamentalDiagramsMultidirectional_2017} &   \cite{_DatabaseJulich_}\\ 
\emph{Unidirectional II} & Exp. & Uni-dir. & - &  Tokyo, Jap, 2018 &  \cite{feliciani_UniversalFunctionCapacity_2018} &  \cite{feliciani_UniversalFunctionCapacity_2018}\\ 
\emph{Bidirectional, steady-state} & Exp. & Bi-dir. & Density &  J\"ulich, Ger, 2013 & \cite{cao_FundamentalDiagramsMultidirectional_2017} &   \cite{_DatabaseJulich_}\\ 
\emph{Bidirectional, pre lane formation} & Exp. & Bi-dir. & - &  Tokyo, Jap, 2020 & \cite{murakami_MutualAnticipationCan_2021}  &  \cite{murakami_MutualAnticipationCan_2021}\\ 
\emph{Zara (Outdoor)} & Obs. & Bi-dir. & - &  Nicosia, Cy, 2007 & \cite{lerner_CrowdsExample_2007a} &  \cite{_Observations_}\\ 
\emph{EWAP (Outdoor)} & Obs. & Bi-dir. & - &  Z\"urich, Swi, 2007 & \cite{pellegrini_YouLlNever_2009} &  \cite{_ObservationsII_}\\ 
\emph{Cross} & Exp. & Multi-dir. & Density &  J\"ulich, Ger, 2013 & \cite{cao_FundamentalDiagramsMultidirectional_2017} &   \cite{_DatabaseJulich_}\\ 
\emph{Antipodal} & Exp. & Multi-dir. & - &  Beijing, PRC, 2019 & \cite{xiao_InvestigationPedestrianDynamics_2019} &  -\\ 
\emph{Students (Outdoor)} & Obs. & Multi-dir. & - &  Tel Aviv, Isr, 2007 & \cite{lerner_CrowdsExample_2007a} &  \cite{_Observations_}\\ 
\emph{Bottleneck} & Exp. & Bottleneck & Corr. Width &  J\"ulich, Ger, 2018 & \cite{adrian_CrowdsFrontBottlenecks_2020} &   \cite{_DatabaseJulich_}\\ 
\addlinespace 
\bottomrule
\end{tabular}
\caption{Summary of the different datasets used in our analysis.}
\label{tab:datasets}
\end{threeparttable}
\end{table} 
\endgroup


Tab.~\ref{tab:datasets} provides a summary of the datasets that have been used, the corresponding references, and, whenever available, download links.
Further details about the experimental setups can be found in the references.

Note that, for the \emph{Waiting} scenario, we had to extract trajectories by ourselves, from existing videos of a controlled experiment conducted within the \emph{BasiGo} project in $2013$ in Germany.
In this experiment, the (27 to 600) participants enter a square area through four entrances.
This area is delimited by crowd control barriers and the crowd is filmed from above.
After the crowd has entered, the participants wait for approximately one minute before egressing through the entrances.
We have extracted parts of this waiting period using the semi-automatic tracking mode of the \emph{PeTrack} software \cite{boltes_AutomaticExtractionPedestrian_2010}.
The corresponding trajectory data is included as supplemental material.

\subsection{Splitting and merging scenarios}
\label{sub:app_A_splitting}

Often, experimental scenarios display different stages and different regions in space. In some cases, specified below, we focused only on part of these stages and regions or we split the data into distinct scenarios.

In most experiments, there is a clear start and end, where the participants start to fill, or begin to leave the measurement area.
We will start reporting those experiments, where we analyzed the pseudo-stationary state and overlooked the beginning and end of each run.
The \emph{Single-File} trajectories were treated as purely one-dimensional by neglecting the transverse coordinate.
The dataset was split into a \emph{dense} and a \emph{sparse} part, where the $6$ runs with the lowest global density are considered as sparse and the rest as dense.
In \emph{Unidirectional I}, we have only used the runs where the width of the entrance and exit corresponds to the total width of the corridor; the other runs rather resemble a bottleneck. 
As for \emph{Unidirectional II}, we have used the totally asymmetric runs where all people walk from one side to the other. 
As there is only a small crowd that passes the measurement area, no steady-state can be analyzed.
For simplicity, we have merged the two unidirectional datasets into a \emph{Unidirectional} scenario.
In \emph{Bidirectional steady-state}, we have used variant $B$, where the participants are instructed to use a fixed exit, i.e. either on the left or on the right at the end of the corridor.
For the \emph{Cross} scenario, we have limited our analysis to the area of the crossing itself.
In particular, we have neglected the corridors leading to the crossing area. 
We have used the variant $A$ of the $90^\circ$ crossing, where people enter from all $4$ sides, without an obstacle in the centre of the crossing. 
Note that, the runs $6$ and $8$, i.e. those with the highest intrusion, where cancelled after some time as the experimenters were afraid that participants could get hurt due to the heavy congestion.
We split the \emph{Cross} dataset in a \emph{sparse} part consisting of the $3$ runs with the lowest global density and the \emph{dense} part with the rest.
In \emph{Bottleneck}, only the runs with a number of participants $N > 40$ were used.
Furthermore we have only used the runs with a high motivation. 
We restricted the analysis to the area right in front of the bottleneck.

In some experiments, we were interested in transient states, i.e. a specific temporal part of the experiment. 
In \emph{Bidirectional pre lane formation}, we have used the runs without any distraction by cellphones, i.e. the baseline condition.
We start the measurement after people have entered the measurement area and end it before the lanes have formed.
For the \emph{Antipodal} experiment, only the first $\approx 2$~$s$ of each run are considered.
In particular, the part before the pedestrians get close to each other.
The runs with a radius of $r = 5$~$m$ and $8$ participants were used.

\subsection{The case of the \emph{Outdoor} scenario (passive observations)}

The \emph{Outdoor} scenario gathers real-world observations from different datasets. 
The complete sequences have been used.
Regarding the \emph{Zara} and \emph{Students} datasets, the data are published only in pixel positions and some frames are missing.
We have, therefore, used the amended data by \cite{alahi_SocialLSTMHuman_2016}, where real-world positions were estimated and enhanced by linear interpolation between the frames.
Regarding the \emph{EWAP} datasets, filmed from a \emph{Hotel} and at the \emph{ETH} campus, the velocities were given with the positions and frames in two-dimensional real-world units.
In contrast to the controlled experiments, many pairs are present in the \emph{Outdoor} scenario.
This has an effect on the structure of the crowd, as we will show in Appendix~C.
In an atomic vision of the crowd, these pairs (featuring specific `intra-molecular' interactions) must be excluded, for the calculation of $\Av$ and especially $\In$.
To do so, we detect pairs according to a simple rule:
Two pedestrians are assumed form a social group if their mean distance is smaller than $1$~$m$, their maximal distance smaller than $1.5$~$m$, and their mutual presence in the scene lasts at least $2$~$s$.

\subsection{Processing and smoothing of trajectories}
Unless otherwise specified, in all datasets each pedestrian is assigned a unique ID, for which a two dimensional real-world trajectory is obtained at a certain frame-rate. 
Although the trajectories were already expressed in real-world coordinates, they featured oscillations due to head sways and empirical noise, which both affect the calculation of $\In$ and above all $\Av$. To smooth out the head sways, a $4$-th order Butterworth filter with critical frequency $0.5$~Hz was applied to the trajectories.
From these positions and times we calculated the velocities as the distance covered in approximately $1$~$s$.

\subsection{Computation of the $\In$ and $\Av$ numbers and filtering out isolated agents}
Subsequently the time-to-collision (TTC) was computed by assuming that each pedestrian is a disk of diameter $\ell = 0.2$~$m$.
This size was chosen in accordance with \cite{karamouzas_UniversalPowerLaw_2014}, in order to limit the number of measured overlaps between disks, which lead to ill-defined TTC and $\In$ values.
Nonetheless, in the very dense experiments, some overlaps are still observed; 
to mitigate this problem, we set an upper bound $\In _{ij}^{\text{max}} = 400$ and $\Av _{ij}^{\text{max}} = 60$ on all computed $\In$ and $\Av$ numbers.

Besides, despite the segmentations mentioned in Sec.~\ref{sub:app_A_splitting}, some scenarios (particularly the \emph{Outdoor} one) remain heteroclite, with a large number of pedestrians that actually walk in isolation.
Another example is the sparse \emph{Cross} scenario, where we want to focus on the half before solving the conflict in the centre of the crossing.
Therefore, we exclude pedestrians with $\Av_i=0$ in the averaged $\Av$.
Excluding these values narrows the datasets down to the parts where interactions really occur and thus yields a much finer and more robust delineation of the different regimes.
This is further related to the fact that only collisions between the hard-cores are taken into account.
A more sophisticated definition of $\Av$ could capture 'soft' collisions with the private or intimate space, which were not captured with the discontinuous TTC.

\section*{Appendix B: Variations of the Phase Diagram}
\label{appendix_B}
\setcounter{subsection}{0}

\subsection{Variations in the definitions of $\In$ and $\Av$}
As we have conceded in the main text, there is a certain freedom in the choice of the definition of $\Av$ and $\In$.
Here, we investigate to what extent this choice impacts the delineation of different regimes.

Let us start with the definition of the Intrusion number $\In$.
For an agent $i$, $\In$ was defined as the sum of all intrusions over the set $\mathcal{N}_i$ of all close neighbors $j$ of $i$, here delimited by $r_{ij} \leqslant 3 \,r_\text{soc}$.
This additivity is similar to the superposition of forces in Physics.
However, we are not dealing with forces in the Newtonian sense and the validity superposition is not granted. For instance, it was found to be unreasonable at least in some situations \cite{Seyfried19}.
Therefore, one might choose an alternative neighborhood where the intrusion is dominated by its maximum value as
\begin{equation}
    \Tilde{\In}_i = \max _{j \neq i}  \In_{ij}.
    \label{Eq:In_MAX}
\end{equation}
The alternative phase diagram is shown in Fig.~\ref{fig:phase_diagram_variants}~\emph{(a)}.
It is difficult to spot any substantial difference to the original diagram:
the delineation is robust under this change.
In an analogous way, for the Avoidance number, we can replace the 
\rev{maximum with a sum.}
{sum over only the most imminent collision with a sum over all possible collisions, viz., \begin{equation}
    \Tilde{\Av} _i = \sum _{j \neq i} \frac{\tau _0}{\tau _{ij}}.
 \end{equation}}

In that case, the delineation of different regimes gets blurred to a large extent.
On the other hand, if more weight is put on the large Avoidance numbers by defining it symmetrically to the Intrusion number, with $k_A=k_I=2$, viz.,
\begin{equation}
    \Tilde{\Tilde{\Av}} _i = \sum _{j \neq i}\left( \frac{\tau _0}{\tau _{ij}}\right)^2,
 \end{equation}
the delineation is at least partly recovered, see Fig.~\ref{fig:phase_diagram_variants}~\emph{(b)}.

\subsection{Use of the density instead of $\In$}

We chose to base the Intrusion number on distances instead of using the local density.
This is partly justified by the ambiguity in the definition of a local density. 
However, we acknowledge that the averaged $\In$ is still closely related to the density, which certainly is the quantity most commonly used  to classify crowds.

\begin{figure}
\minipage{0.27\linewidth}
    \centering
    \textbf{\emph{(a)}} $\max \left(\In\right)$
    \includegraphics[width= 0.9 \linewidth]{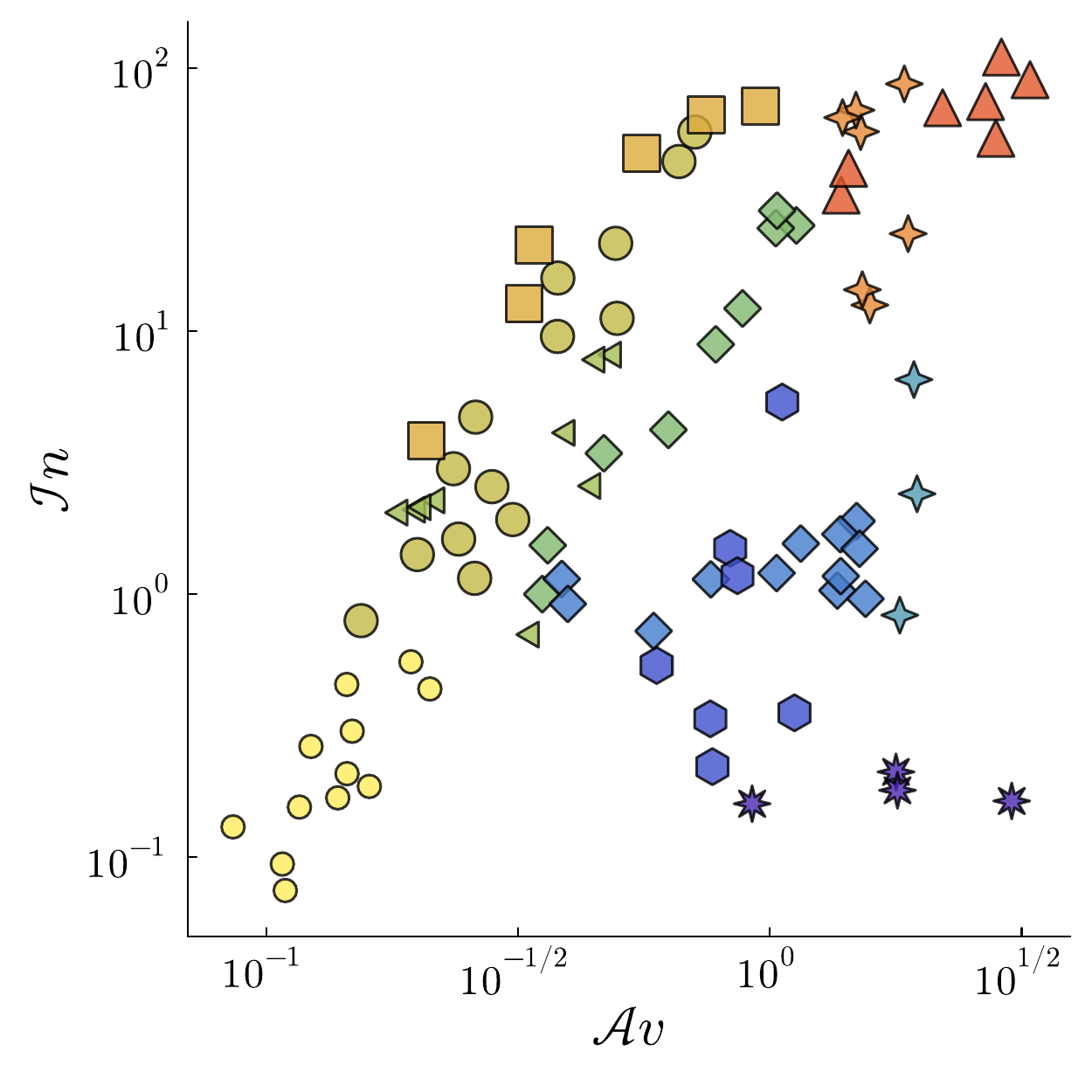}
\endminipage\hfill
\minipage{0.27\linewidth}
    \centering
     \textbf{\emph{(b)}} \emph{$\sum (\tau_0 / \tau) ^2$}
    \includegraphics[width= 0.9 \linewidth]{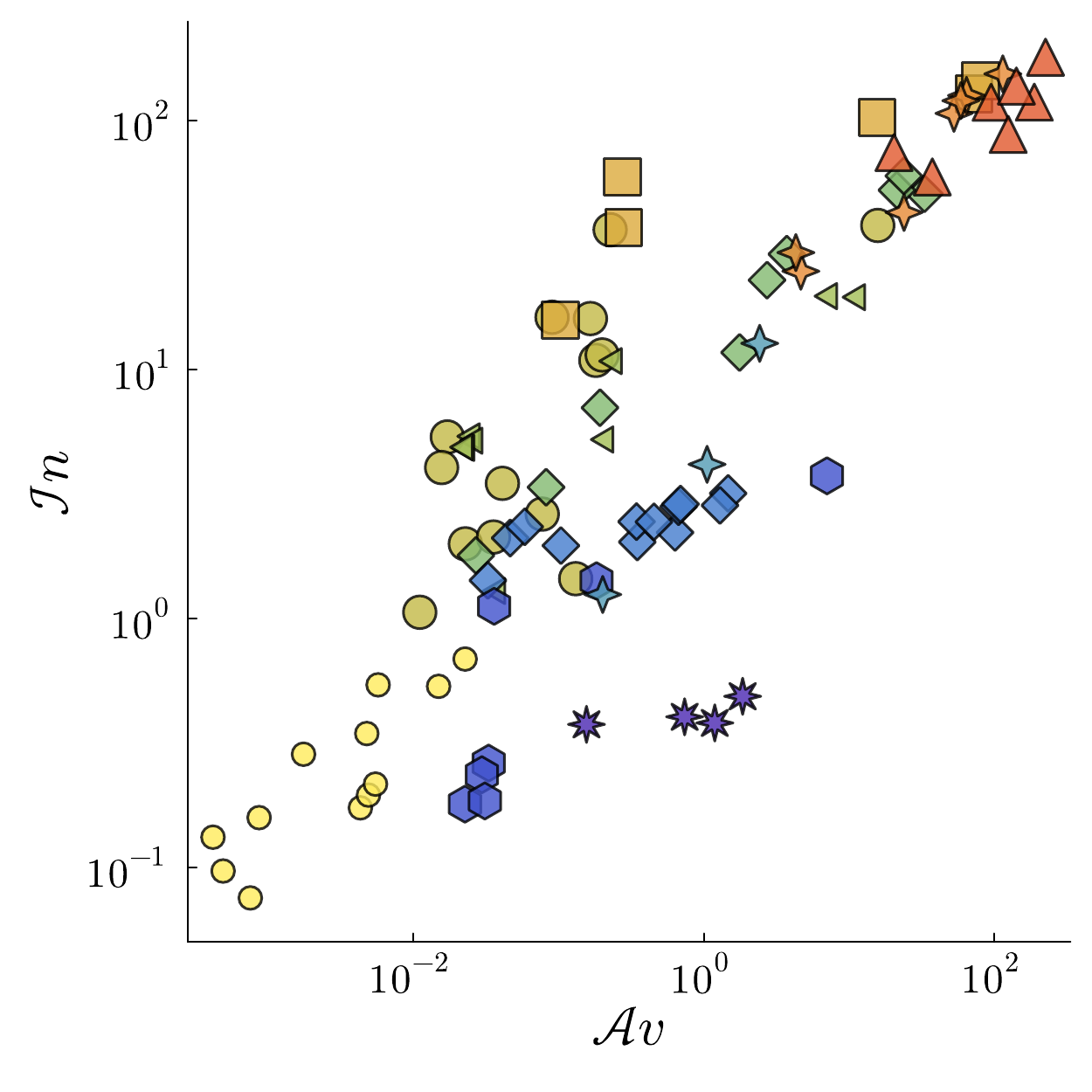}
\endminipage\hfill
\minipage{0.27\linewidth}
    \centering
     \textbf{\emph{(c)}} \emph{Density}
    \includegraphics[width= 0.9 \linewidth]{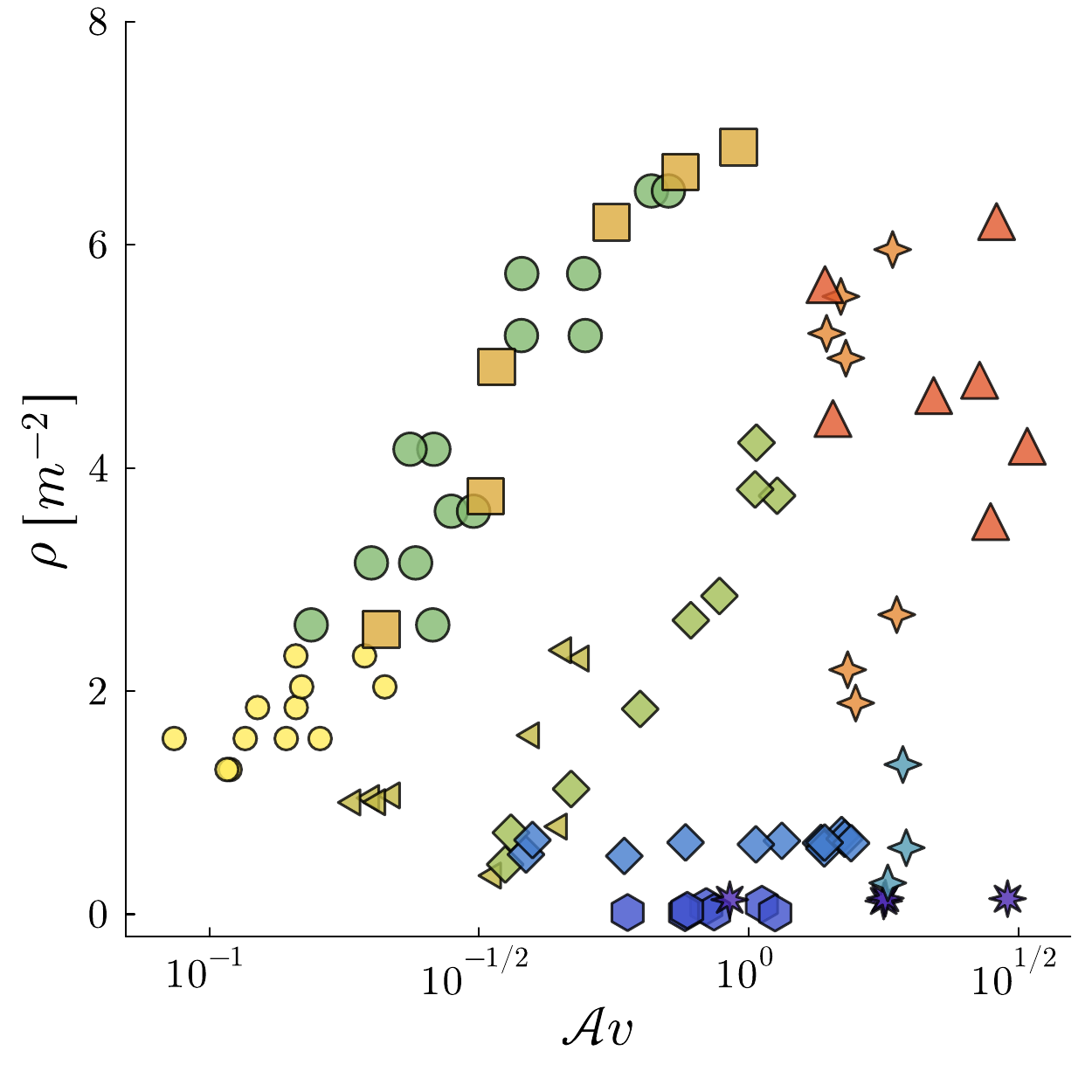}
\endminipage\hfill
\minipage{0.15\linewidth}
    \centering
    \includegraphics[width= 0.9 \linewidth]{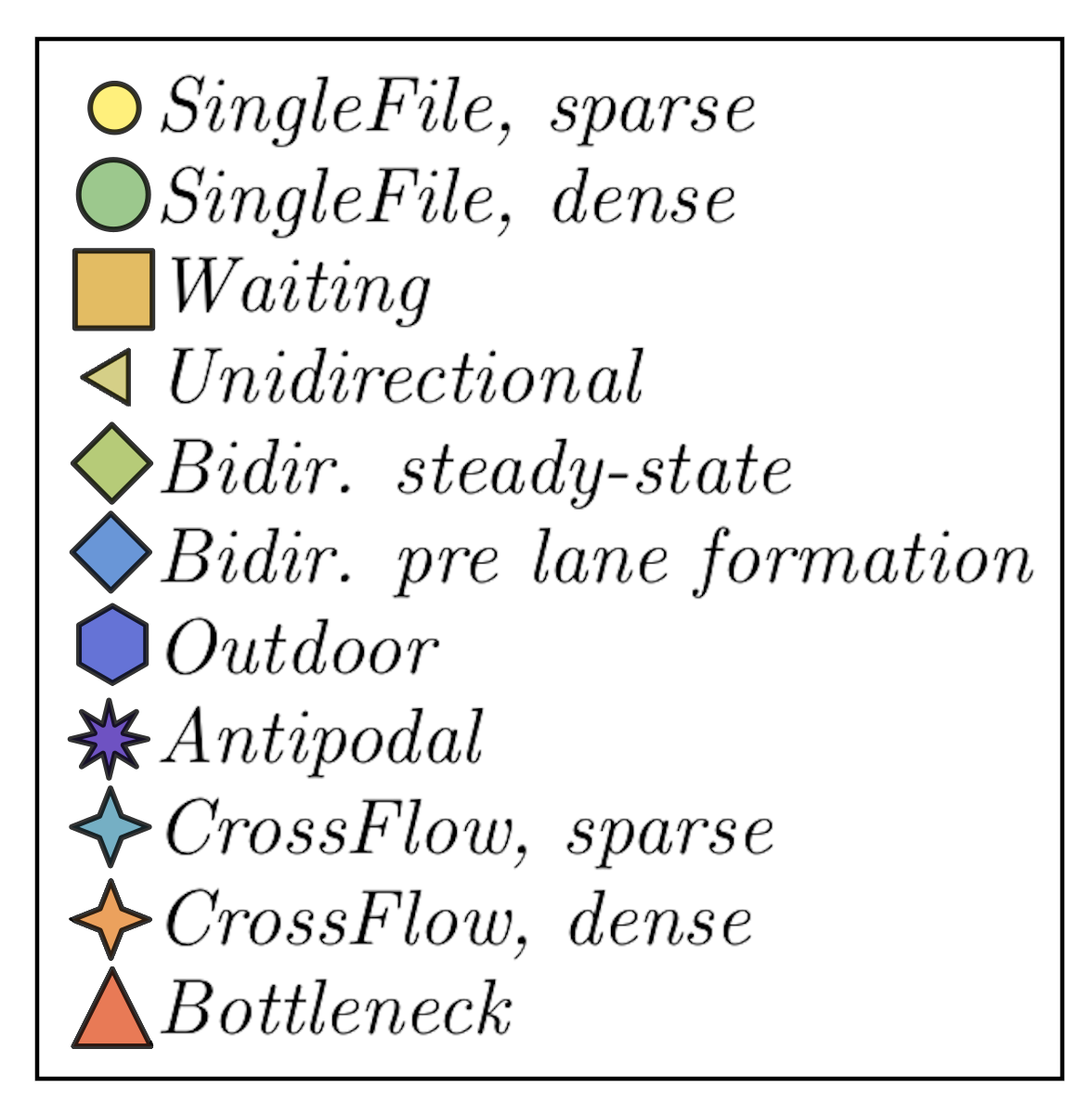}
\endminipage\hfill
\caption{Empirical delineation of different regimes with alternative definitions of the Avoidance number $\Av$ and the Intrusion number $\In$ (or the global density).}
\label{fig:phase_diagram_variants}
\end{figure}

In Fig.~\ref{fig:phase_diagram_variants}~\emph{(c)}, we have substituted the Intrusion number with the global density $\rho$, calculated as the number of pedestrians divided by the available space. 

To enable us to plot the \emph{Single-File} data  along with the rest, the one-dimensional density, calculated as the number of people divided by the length of the track, was rescaled according to \cite{seyfried_FundamentalDiagramPedestrian_2005}, where we assumed a width of $0.3$~$m$.
The delineation of different regimes is still clearly visible.
Only the \emph{Single-File} data strongly deviates from the original diagram.
In particular, relative to the rest of the diagram, moderate intrusions seem to correspond to high densities.
In the \emph{Single-file} scenario, the pedestrians do not have neighbors to the sides.
Therefore, the deviation might actually be reflected in the subjective feeling of the pedestrians.
On the other hand, the deviation could also be explained by the presence of obstacles, e.g. walls, which are very close to all of the agents in the single-file scenario and have not been taken into account.

Besides, the \emph{Cross} scenario and the \emph{Bottleneck} scenario are at lower densities, but higher $\In$, compared to the \emph{Waiting} scenario.
In the latter people are distributed very homogeneously, whereas inhomogeneities are conspicuous in the former, including regions of tight packing. The Intrusion number puts more weight on these.


One might think that, since crowd regimes are typically classified on the basis of the global density, the agent-centred intrusion variable $\In_i$ could harmlessly be substituted by the local density $\rho_i$. However, this is dubious. There is of course the question of the hydrodynamic limit, in which fluids are, but pedestrian crowds are not: there is no strict separation of scales between the pedestrian size and the system size. But this is admittedly only a minor problem:
one can make use of Voronoi cells to define a local density in crowds, with the inverse area of the cell providing an agent-centred local density. Much more annoying is the fact that such Voronoi-based density fails to blow up intrusions  into social `bubbles' or, even more importantly,  biophysical contacts. Indeed, consider a large room in which there are only two individuals; even if one person significantly intrudes upon the other, the associated Voronoi density for each pedestrian remains minimal. This seemingly artificial scenario is typically used in 'stop-distance' experiments in the context of proxemics, to probe people's reactions. 
In these experiments, a subject approaches or is approached by an assistant until the subject begins to feel uncomfortable due to the proximity of the other person \cite{hayduk_PersonalSpaceWhere_1983}.
In contrast, the comfort distance is directly reflected by the presence of the `social radius' $r_{\text{soc}}$ in $\In_i$; furthermore, $\In_i$ also captures the true physical size $\ell_{\text{min}}$, where the quasi-incompressibility of the human body translates into the divergence of $\In_i$ to $+\infty$ as this physical limit is approached.

\subsection{Effect of correlations between the empirical values of $\In$ and $\Av$}

There may be some correlation between $\In$ and $\Av$, so that when one of them changes, the other changes as well.  The systematic increase in $\Av$ in unidirectional flows as the crowd gets denser, even after application of the Butterworth filter to correct head sway, supports this impression. 
In the main text, we have seen that $\In$ and $\Av$ are sufficiently independent to allow for a proper distinction between the typical scenarios encountered in pedestrian streams.
In any case, even some degree of interdependence would mostly result in a skewed diagram (given that the $\In$ and $\Av$ axes have been plotted orthogonally although they should not) with no impact on its topology.

\section*{Appendix C: Structure of Crowds}
\label{appendix_C}
\setcounter{subsection}{0}

We have used the pair-distribution function (pdf) to probe the structure of crowds and, subsequently, to identify the best descriptor of its self-organized structure.
Here, we give more details on the calculation of the pdf. Then we will investigate the structure of crowds in different scenarios in more detail.

\subsection{Definition of the observable}
\begin{figure}
    \centering
    \includegraphics[width= 0.5 \linewidth]{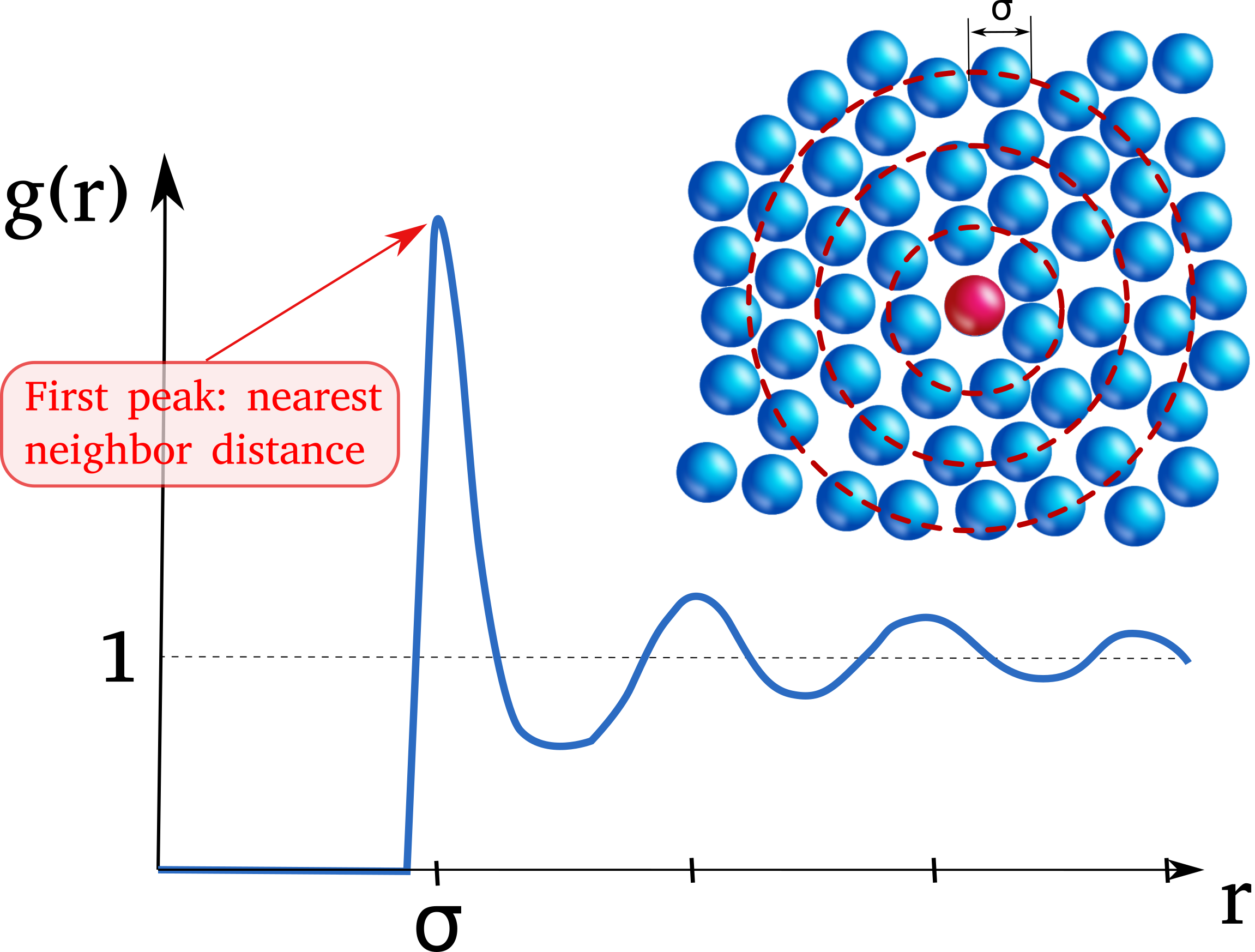}
\caption{Exemplary figure of the pair-distribution function
}
\label{fig:pair_dist_illustrative}
\end{figure}

The pdf is generally used to infer the atomic or molecular structure of materials.
An illustration of the relation of the pdf to the structure of a material is shown in Fig.~\ref{fig:pair_dist_illustrative}; an experimental example of a pdf, measured for solid and liquid Argon, can for example be found in \cite{franchetti_RadialDistributionFunctions_1975}. 
In our case, we calculate the pdf of crowds according to \cite{karamouzas_UniversalPowerLaw_2014}. 
For some variable $x$ the pdf is given by the probability that two pedestrians are separated by $x$ normalized by the probability $P_{\text{NI}}(x)$ that two non-interacting pedestrians are separated by $x$, in particular $g(x) = P(x)/P_{\text{NI}}(x)$. This normalization is aimed at correcting the lack of translational invariance in crowd observations.

While $P(x)$ can be simply estimated by the relative frequencies in the dataset, $P_{\text{NI}}(x)$ is in principle unknown.
However, it can be estimated by randomizing either the spatial or the temporal information.
To estimate the distribution, we used strict binning with bins of size $0.1$~$m$ or $0.1$~$s$.

\subsection{Interpretation in various scenarios}

This procedure is best understood in the case of single-file motion.
Here, the observational area is limited to the $x$-coordinate and ranges from $0$ to $L$.
As all pedestrians enter the scene on the left and leave on the right, all positions are equally likely.
However, due to the limited size of the area, finite-size effects strongly suppress large distances.
Therefore we can estimate $P_{\text{NI}}(x)$ by calculating the distribution of distances between points that are randomly positioned on the interval $[0, L]$.
Another way to estimate $P_{\text{NI}}(x)$ is by randomzing the time-information, i.e. by creating a 'time-scrambled' version of the dataset as proposed by \cite{karamouzas_UniversalPowerLaw_2014}.
This ensures that the distances calculated in the scrambled dataset correspond to non-interacting pedestrians as they have not been in the same frame originally.
Both procedures lead to the same result in the case of single-file motion.

In the case of the \emph{static} crowd, the scrambling of temporal information cannot be employed, as the pedestrians are hardly moving.
Therefore, we assume that all positions within the rectangle are equally likely.
As we neglect the edges of the observational area, where people  lean on the crowd-control barrier, this assumption is justified.

In the \emph{Outdoor} dataset, alongside finite-size effects, one has to account for different forbidden areas (like trees or cars) in the middle of the scenes and the different areas where people tend to enter or leave the scene. 
This is achived by randomizing the time-information.


\begin{figure}
\minipage{0.33\linewidth}
    \centering
     \textbf{\emph{(a)}} \emph{Corridor}
    \includegraphics[width= 0.99 \linewidth]{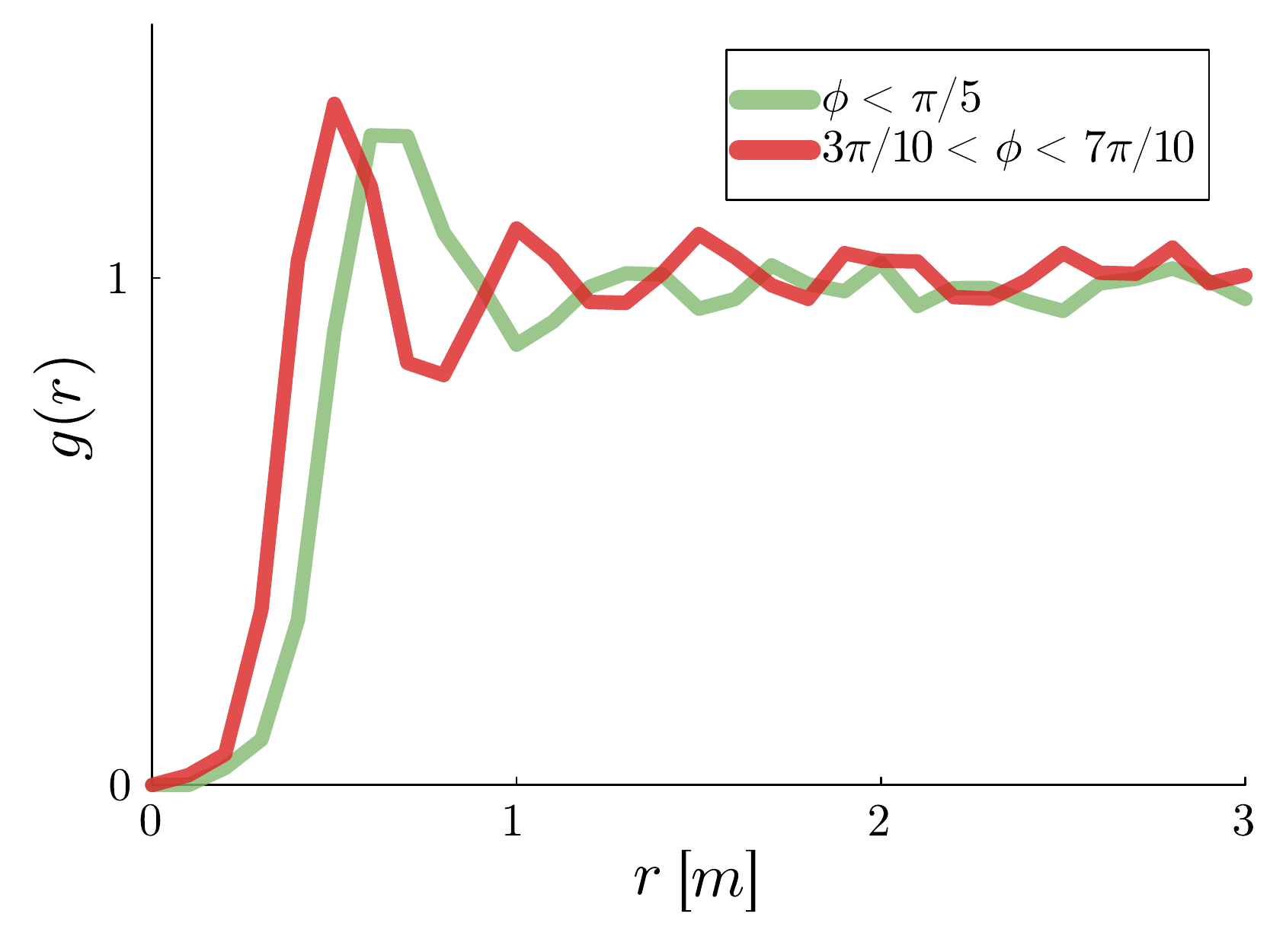}
\endminipage\hfill
\minipage{0.33\linewidth}
    \vspace{0.05\linewidth}
    \centering
     \textbf{\emph{(b)}} \emph{Outdoor}
    \includegraphics[width= 0.99 \linewidth]{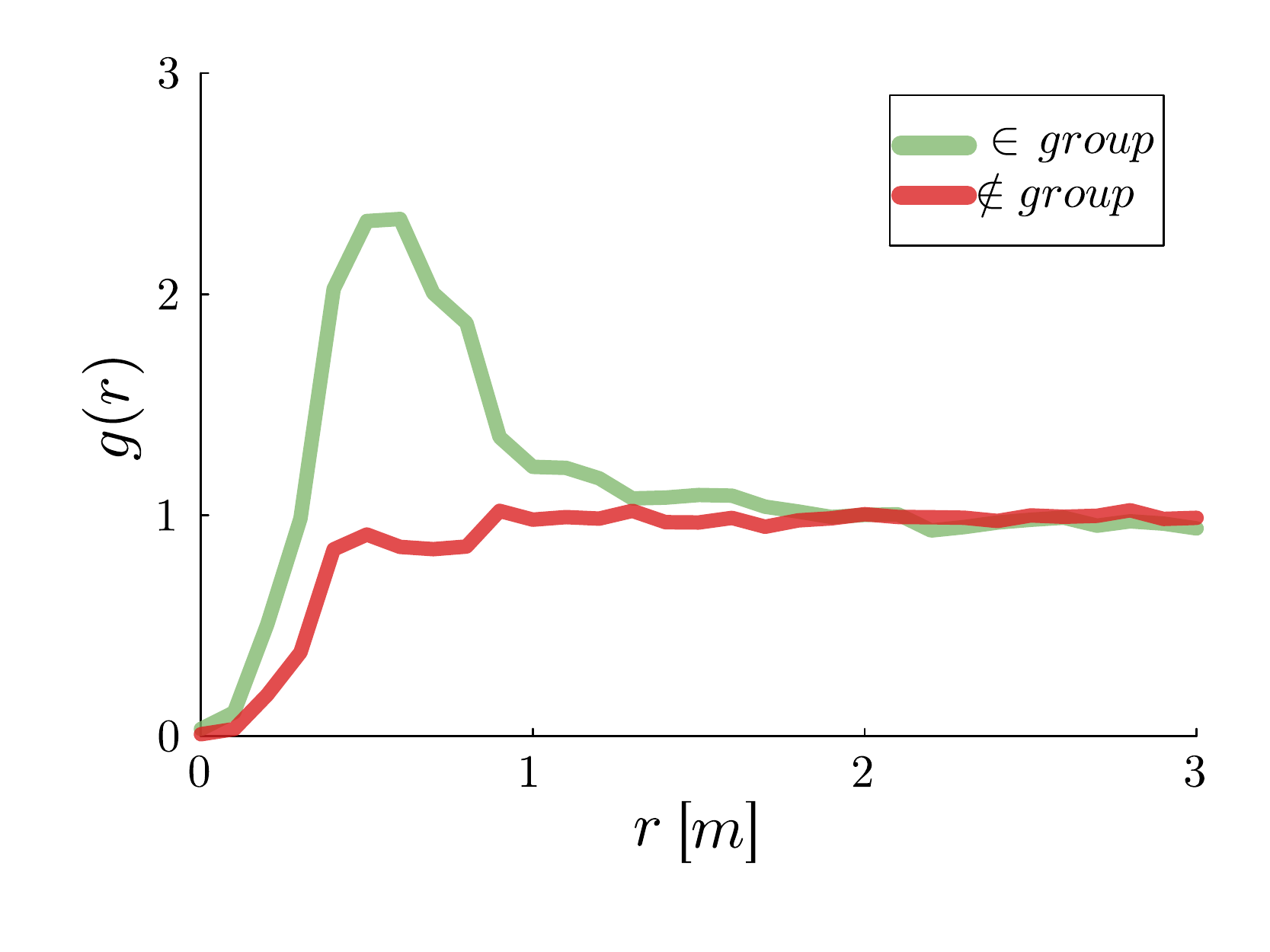}
\endminipage\hfill
\minipage{0.33\linewidth}
    \vspace{0.05\linewidth}
    \centering
     \textbf{\emph{(c)}} \emph{Outdoor}
    \hfill
    \includegraphics[width= 0.99 \linewidth]{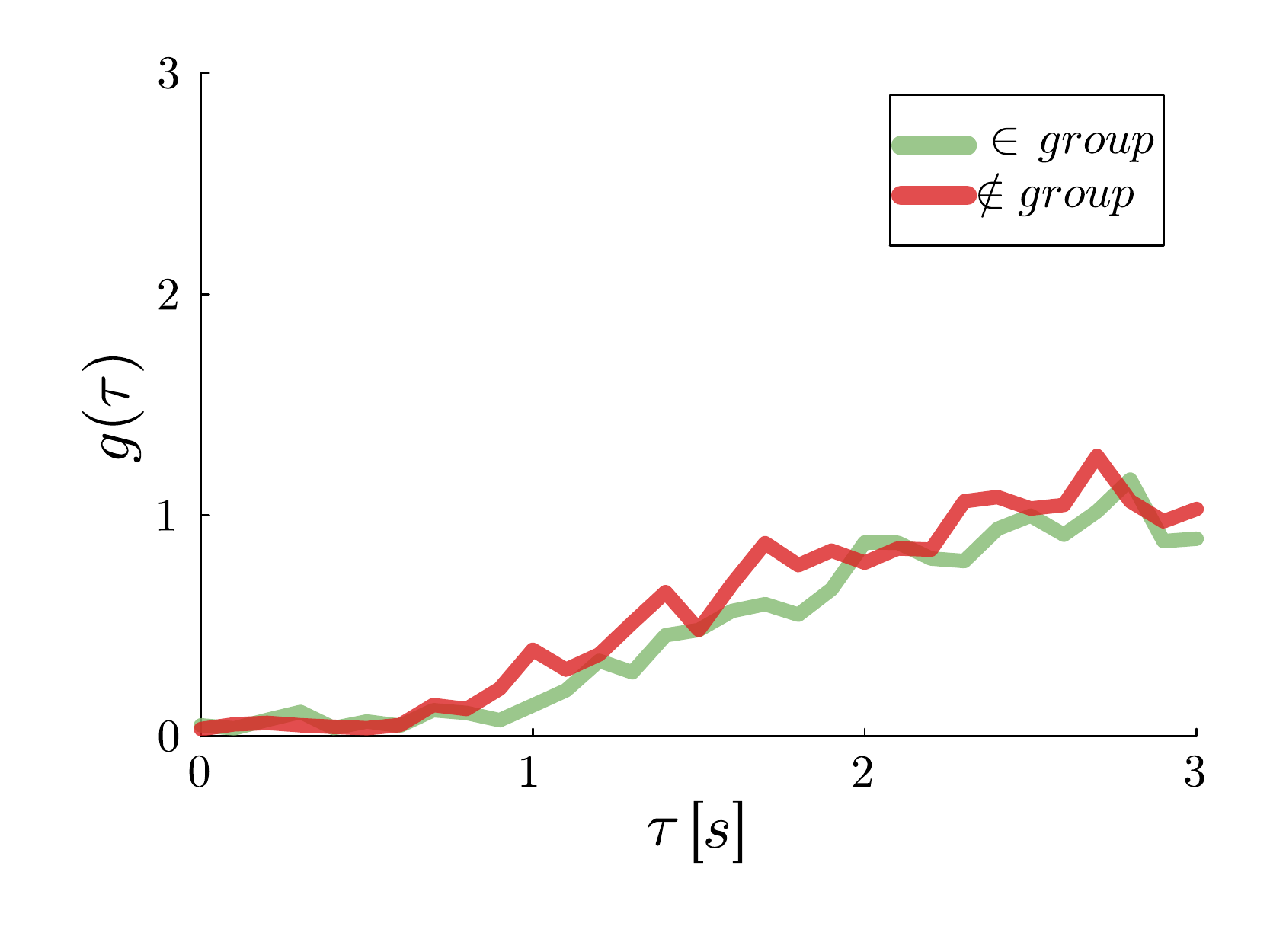}
\endminipage\hfill

\caption{Evaluation of the pdf $g(r)$ in different scenarios. In \emph{(a)}, i.e. the \emph{Unidirectional I} dataset, the run with $\rho = 2.1/m^2$ the position of the peaks depends on the angle $\phi$. In particular, the longitudal distances (to the front) are larger than the transversal distances (to the sides). In \emph{(b)}, the \emph{Outdoor} dataset, the marked peak at small distances to the sides arises from many pairs which are present in the observation. This cannot be observed in the TTC-based pdf $g(\tau)$, cf. \emph{(c)}.}
\label{fig:pair_dist_structure}
\end{figure}


\subsection{Insight given by the pdf into the crowd's structure in different scenarios}

Probably the most confined and homogeneous pedestrian experiments were conducted by \cite{seyfried_PhaseCoexistenceCongested_2010}: high density, periodic, single-file experiments with soldiers as the participants.
The corresponding pdf is shown in Fig.~1 \emph{(C)} in the main text for a single run at $\rho = 2.1/m$,
where  long-ranged correlations in the pdf can be seen, owing to
the strong homogeneity, combined with the spatially confined setting.
The pdf displays peaks that are well separated and located at the integral multiples of the mean spacing, i.e. correspond to the $k$-th neighbor.

Let us now turn to the unidirectional flow through a corridor with open boundaries at a density of $\rho = 2.1/m^2$. 
The corresponding pdf is shown in Fig.~\ref{fig:pair_dist_structure}~\emph{(a)}. 
Multiple peaks are still visible but the correlation length is much smaller.
This time the curves are binned according to the angle $\phi$ between the velocity $\bs{v}_i$ and the vector connecting the centers of two pedestrians $\bs{r}_{ij}$.
The two curves exhibit a difference between longitudinal (\emph{green}) and transversal (\emph{red}) distances, reflecting the existence of anisotropy.
In particular, pedestrians keep smaller distances to their sides than to the front.
The angular dependence may have practical implications as to whether the capacity of a corridor increases linearly or step-wise with its width.

\subsection{Effect of pairs and social groups on the pdf}
Finally, we have calculated the pdf for the \emph{Outdoor} dataset.
As earlier, we have identified pairs according to a simple classification. 
Very distinct curves are seen in Fig.~\ref{fig:pair_dist_structure}~\emph{(b)}, depending on whether pedestrian $i$ is part of a social group ($\in group$) or not ($\notin group$).
Pedestrians that form a social group want to interact (e.g. talk) and, therefore, stay in each other's personal space, generally walking abreast.
This attractive interaction leads to a strong peak at small (transversal) distances.
Apart from this and a strong short-ranged repulsion, no spatial structure can be seen.
The pdf bears resemblance to a mixed gas consisting of single atoms and molecules.
If the pdf is calculated for the time-to-collision, cf. Fig.~\ref{fig:pair_dist_structure}~\emph{(c)}, the two curves collapse onto each other, because proximity in space is not associated with a risk of imminent collision: Pair members want to stay relatively close to each other but do not want to collide.
The main text has underlined that the finding of \cite{karamouzas_UniversalPowerLaw_2014} (namely, that the TTC is a more suitable descriptor than the spatial distances) is valid in a certain regime only (i.e. low $\In$, moderate $\Av$).

By turning the original argument upside down, we contend that further restrictions are necessary when it comes to social groups: distance-based (`proxemic') interactions within each group are combined with TTC-based (avoidance) interactions with other people.
These two levels are reflected in their corresponding pdf:
 the peak at short distances in $g(r)$ (but not in $g(\tau)$) is only present for members of social groups.
Incidentally, this also explains the large variations in the pdf $g(r)$ in \cite{karamouzas_UniversalPowerLaw_2014}, binned by rates of approach: in \cite{karamouzas_UniversalPowerLaw_2014}, pairs were not excluded and the rate of approach of their members is very small, and thus falls in one specific bin.


\begin{figure}
    \centering
        \includegraphics[width= 0.33 \linewidth]{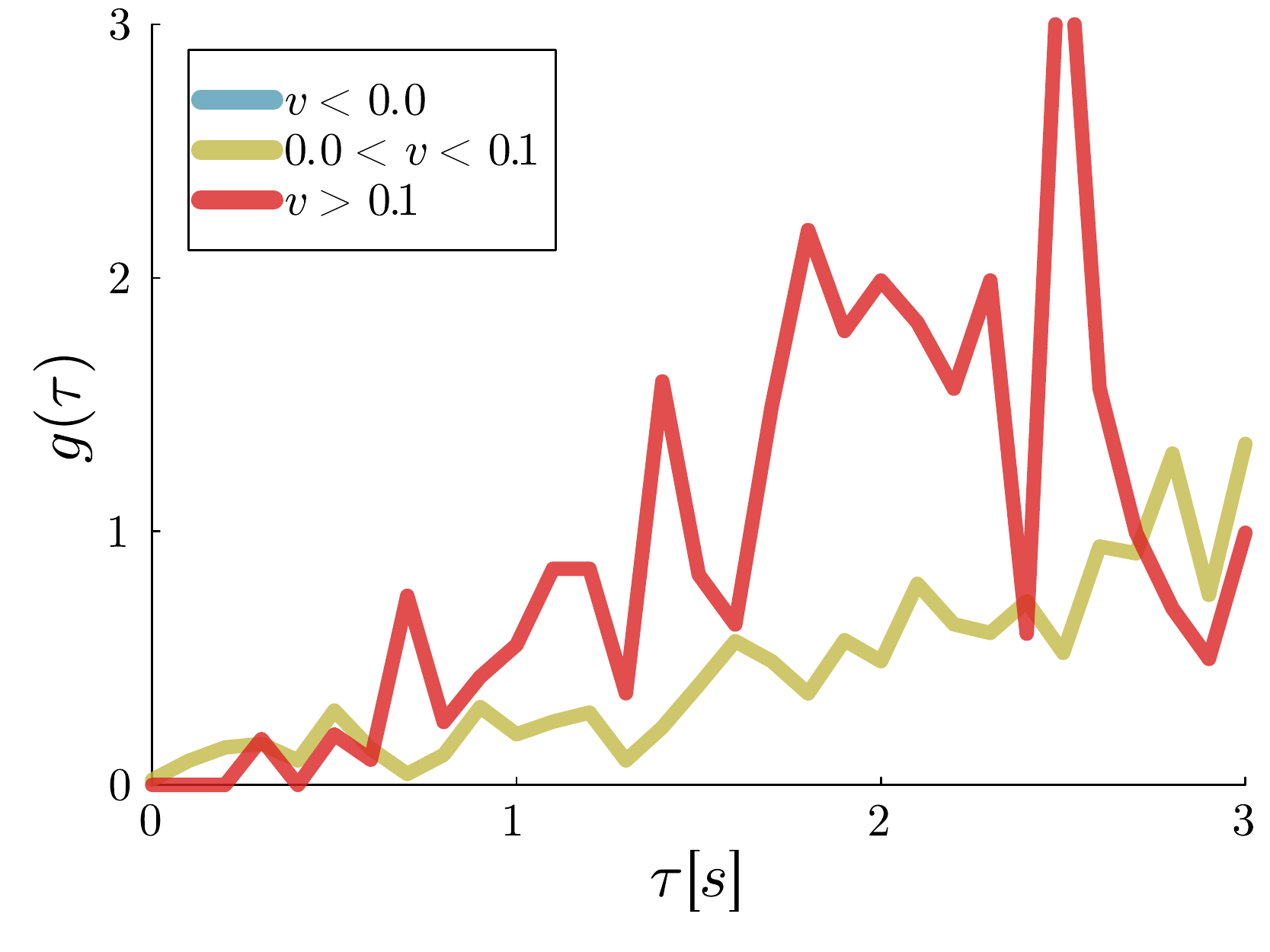}
\caption{The pair-distribution function $g(\tau)$ for the \emph{Waiting} dataset binned according to different rates of approach $v$.}
\label{fig:waiting_ttc_binned}
\end{figure}


\subsection{\emph{Waiting} scenario}

In Fig.~\ref{fig:waiting_ttc_binned} we show that for the \emph{Waiting} dataset the curves of $g(\tau)$ do not collapse onto each other if binned according to the rate of approach $v$.
For negative rates of approach (hence, infinite TTC), no curve can be plotted in Fig.~\ref{fig:waiting_ttc_binned} even though the structure is independent of it in real space (Fig.~1~\emph{(D)}, in the main text).

The failure of $g(\tau)$ to describe the dynamics in the high $\In$ and low $\Av$ regime has bearing on the modeling of starting waves, see e.g. \cite{tomoeda_PropagationSpeedStarting_2012, rogsch_StartWavesPedestrian_2010a}.
Indeed, whether the pedestrians ahead of a reference agent stand still ($v \approx 0$) or move ahead ($v<0$), their TTC is infinite, so that a purely TTC (or $\Av$)-based model will not reproduce the backward-propagating starting
wave that is observed in reality \cite{bain_DynamicResponseHydrodynamics_2019}.

\section*{Appendix D: Perturbative Analysis}
\label{appendix_D}
\setcounter{subsection}{0}

In this Appendix, we detail how a generic perturbative expansion of the cost function $\mathcal{C}$ 
gives rise to the proposed $\In$, $\Av$ and $\Av*\In$-models.

\subsection{Generic cost function}
Consider $N$ circular agents of diameter $\ell$.
The position of agent $i$ is $\bs{r}_i (t) = \left( x_i(t), y_i(t)\right)$, $\bs{v}_i = \dot{\bs{r}}_i(t)$ the velocity, $\bs{a}_i =  \ddot{\bs{r}}_i(t)$ the acceleration, and $\bs{v}_{\text{des}, i}$ the desired velocity.
Let $\mathcal{R}(t)=\left(\bs{r}_1(t),\dots,\bs{r}_N(t)\right)$ denote the set of all positions at time $t$ and $\mathcal{V}(t)=\left(\bs{v}_1(t),\dots,\bs{v}_N(t)\right)$ the set of all velocities.

We have argued that, in most scenarios, the way in which pedestrians choose their velocity is strongly influenced 
by the Intrusion and Avoidance variables, on top of the agent's desire. Accordingly, it can be approximated by
\begin{equation}
    \bs{v}_i^\star = \underset{ \bs{v} \in \mathbb{R}^2 }{\text{arg min}}\,\mathcal{C}_i\left[\bs{v},\In_i(\bs{r}), \Av_i \left( \bs{v}\right) \right],
    \label{eq:dynamics2}
\end{equation}
where we recall that $\bs{v}$ denotes the test-velocity and $\bs{r} = \bs{r}_i (t) + \bs{v} \delta t$ the associated test position. We have shortened the dependencies $\In_i(\bs{r}) = \In_i\left(\bs{r}_i (t)+ \bs{v} \delta t, \mathcal{R}(t)\right)$
and $\Av_i \left( \bs{v}\right) = \Av_i \left( \bs{v}, \mathcal{R}(t), \mathcal{V}(t)\right)$; this means that the TTC $\tau_{ij}$ entering $\Av_i$ are evaluated using the current positions and velocities of all agents, except that
agent $i$'s velocity is substituted by $\bs{v}$.

Taking advantage of the simple form of Eq.~\ref{eq:dynamics2} and without any other major assumption on the microscopic dynamics, we will show that expanding this equation naturally gives rise to simple asymptotic models.

\subsection{Reference situation: The isolated agent}

The reference situation is that of the isolated pedestrian, walking at their desired velocity $\bs{v}_{\text{des}, i}$. (The positional dependence of the desired velocity is not explicitly written.)
In the absence of interactions, $\In_i = 0$ and $\Av_i = 0$,
and Eq.~\ref{eq:dynamics2} reduces to
\begin{equation}
\bs{v}_i^\star = \underset{ \bs{v} \in \mathbb{R}^2 }{\arg \min}\, C_i \left(\bs{v},0,0\right),
\end{equation}
where the minimum should be reached for $ \bs{v}_i^\star = \bs{v}_{\text{des}, i}$. It follows that
\begin{equation}
\bs{0} = \nabla _{v} \, \mathcal{C}_i^{(0)}\left(\bs{v}\right) \Big|_{\bs{v}=  \bs{v}_{\text{des},i}},
\end{equation}
where $\nabla _{v} = \left( \partial / \partial v_{x_i}, \partial / \partial v_{y_i}\right)$, and the Hessian matrix  $\underline{M} = \partial ^2 C_i^{(0)}\left(\bs{v}\right)/\partial \bs{v}^2 \vert _{\bs{v}=  \bs{v}_{\text{des}, i}}$ is positive definite.
In the following we will make an assumption of isotropy around the optimal velocity, in which case $\underline{M}$  is an identity matrix multiplied by a positive scalar $\gamma_i$. Since the cost function can be arbitrarily rescaled, one can set $\gamma_i$ to unity without loss of generality. 
Then, up to second order, the cost-function for an isolated pedestrian is
\begin{equation}
\mathcal{C}_i^{(0)}\left(\bs{v}\right) = \left(\bs{v}_{\text{des}, i} - \bs{v}\right) ^2.
\label{eq:C_0_isotropic}
\end{equation}
The cost increases as the squared Euclidean distance between the test velocity and the desired velocity. 
In particular, deviations in the magnitude and the direction of the desired velocity are similarly penalized.
While a common assumption the literature, this need not be exact in reality.

\subsection{The $\Av$-model}

Let us now perturb the non-interacting system by letting $\Av_i \neq 0$, while maintaining  $\In_i = 0$ for the time being, considering for example two joggers that are still well separated but face an anticipated collision.
In this case, the cost function can be expanded into
\begin{equation}
\mathcal{C}_i \left[ \bs{v}, 0, \Av_i\left(\bs{v}\right) \right] \approx  \left(\bs{v}_{\text{des}, i} - \bs{v}\right) ^2 + \alpha_i \Av_i\left(\bs{v}\right),
\label{eq:AV_Model}
\end{equation}
Here, we introduced $\alpha _i := (\partial \mathcal{C}_i / \partial \Av_i )_{\Av_i=0}$, which is non-negative because pedestrians shun collisions.
Complemented with relaxation process, where the actual velocity relaxes towards $\bs{v}_i^\star$ as $\bs{\dot{v}}_i = \left(\bs{v}_i^\star - \bs{v}_i(t)\right)/\tau _R$, we recover the proposed $\Av$-model.

At this stage, it is worth underlining that this expansion holds for any dependence of $\Av_i$ on the TTC $\tau_{ij}$; in particular, no obvious symmetry prescribes the value of the exponent $k_A$ entering the definition of  $\Av_i$. Should one take $k_A=1$, as we did for the simulations, the resulting $\Av$-model shares the same dependence on the TTC as the RVO model \cite{vandenberg_ReciprocalVelocityObstacles_2008}, whereas setting $k_A=2$ is more consistent with the empirical analysis of \cite{karamouzas_UniversalPowerLaw_2014}. Qualitatively, the $\Av$-models obtained for different $k_A$ will display comparably, but the degree to which the agents will anticipate collisions and swerve from their straight paths will vary.

\subsection{The $\In$-model}

Now, we assume $\In_i \neq 0$ while $\Av_i = 0$, for example in a moderately dense, static crowd.
Perturbatively,
\begin{equation}
\mathcal{C}_i \left[ \bs{v}, \In_i \left( \bs{r}\right) \right] \approx \left(\bs{v}_{\text{des}, i} - \bs{v}\right) ^2 + \tilde{\beta}_i \, \In_i \left( \bs{r}\right),
\label{eq:IN_expansion}
\end{equation}
where $\tilde{\beta} _i := (\partial \mathcal{C}_i / \partial \In_i )_{\In_i=0}$.
This function is extremal for $\bs{v}=\bs{v}_i^\star$, which implies that
\begin{equation}
\bs{0} =   \nabla _{\bs{v}} \left(\bs{v}_{\text{des}, i} - \bs{v}\right) ^2 \big\vert_{\bs{v}=\bs{v}_i^\star} + \tilde{\beta}_i \, \nabla_{\bs{v}} \In_i\left( \bs{r} \right)\big\vert_{\bs{r}=\bs{r}_i(t)+\bs{v}_i^\star \delta t}.
\end{equation}
Therefore, by substitution,
\begin{equation}
\bs{0} = -2 \left(\bs{v}_{\text{des}, i} - \bs{v}_i^\star \right) + \beta_i \delta t \,  \nabla_{\bs{r}} \In_i\left( \bs{r} \right)\big\vert_{\bs{r}=\bs{r}_i(t)+\bs{v}_i^\star \delta t},
\end{equation}
where $\nabla _r  = (\partial / \partial x, \partial / \partial y)$, if $\bs{r}=(x,y)$.
For sufficiently small $\delta t$, one can assume $\bs{r} \approx \bs{r}_i(t)$ and subsequently obtain
\begin{equation}
\bs{v}_i^\star =  \bs{v}_{\text{des}, i} - \beta_i\,  \nabla _r \In_i\left( \bs{r}_i(t) \right),
\label{eq:In_Model}
\end{equation}
where $\beta _i := \tilde{\beta}_i \delta t / 2$.
\rev{
Eq.~\ref{eq:In_Model}, if combined with a relaxation time-scale, the proposed $\In$-model.
}
{
Combined with a relaxation time-scale, Eq.~\ref{eq:In_Model} is the proposed $\In$-model.
}

\subsection{The $\Av*\In$-model}

Let us now turn to the general case $\Av_i \neq 0$ and $\In_i \neq 0$.
Combining the expansions of Eqs.~\ref{eq:AV_Model} and \ref{eq:IN_expansion}, and expressing the extremal condition for $\bs{v}_i^\star$ (Eq.~\ref{eq:dynamics2}), we arrive at
\begin{equation}
\mathcal{C}_i \left[ \bs{v}, \In_i \left( \bs{r}\right) ,\Av_i\left(\bs{v}\right)\right] 
=
 \underset{ \mathcal{C}^{(\In)}_i }{ \underbrace{     \left(\bs{v}_{\text{des}, i} - \bs{v}\right) ^2  + \tilde{\beta}_i \, \In_i\left( \bs{r} \right) }}  + \alpha_i \,  \Av_i\left( \bs{v} \right).
\label{eq:pert_InAv_1}
\end{equation}
Given that $\Av_i \neq 0$, the solution $\bs{v}_i^{(\In)}:= \bs{v}_{\text{des}, i} - \beta_i \,  \nabla _{\bs{r}} \In_i\left( \bs{r}_i(t) \right)$ minimizes $\mathcal{C}^{(\In)}_i$ (or, in other words, the $\In$-model), to leading order, $\mathcal{C}^{(\In)}_i$ is well approximated by the parabola
\begin{equation}
\mathcal{C}^{(\In)}_i \approx  \left[\bs{v}_{\text{des}, i}   - \beta_i \,  \nabla _{\bs{r}} \In_i\left( \bs{r}_i(t) \right) - \bs{v} \right]^2
\end{equation}

It follows that
\begin{equation}
    C_i \left[ \bs{v}, \In_i \left( \bs{r}\right) ,\Av_i\left(\bs{v}\right)\right] \approx \left[\bs{v}_{\text{des}, i}   - \beta_i \,  \nabla _{\bs{r}} \In_i\left( \bs{r}_i(t) \right) - \bs{v} \right] ^2 + \alpha _i\, \Av_i\left( \bs{v} \right),
\end{equation}
which is the cost function of the $\Av*\In$-model.

\section*{Appendix E: Connection between the dynamics and the structure}
\label{appendix_E}
\setcounter{subsection}{0}

In the main text, we have delineated asymptotic regimes ($\Av \to 0$ or $\In \to 0$) in the dynamics of crowd flows, in which the microscopic dynamics of each agent are governed by only the  intrusion variable $\In_i$ or the avoidance variable $\Av_i$. 
Empirical evidence has revealed distinct types of crowd arrangement in these two regimes, characterised by the pdf $g(r)$ and $g(\tau)$, respectively.

Theoretically, it is thus tempting to prove that, if, say, the dynamics hinge on the distance-based variable $\In_i$, independently of the TTC, then the crowd's structure will be characterised by a pdf $g(r)$ that is independent of other variables such as the rate of approach $v$. While this is of course true if an equilibrium state is reached in which
the separations between particles are a function of a potential that only depends on the intrusion variable (hence $r$), it so happens that it does not hold systematically, because $r$ and $v$ are not independent variables.

Let us illustrate this with a minimal example. Consider two `sticky' agents moving on a periodic line and interacting with a distance (i.e., $\In$)-based criterion. More precisely, at each time step, they randomly select a velocity $v_i = \pm 1$, unless they are separated by less than a short distance $r_0$, in which case they come to a halt ($v_i=0$). As time moves on and random encounters occur, more and more particles will stick together in configurations $r<r_0$ and $v=0$. Thus, even if initially the separations were independent of the rate of approach $v$ (binned into two bins, $v=0$ or $v\neq 0$), configurations with $v=0$ will be more and more biased towards short separations, compared to a non-interacting situation. It follows that
the pdf $g(r)$ may depend on the rate of approach, even if the dynamics are governed by a purely distance-based criterion.

\section*{Appendix F: Description of the movie provided as Supplemental Material}
\label{appendix_F}
The supplementary movie compares the dynamics of pedestrians observed in controlled experiments with the predictions of the asymptotic "Av", "In", and "Av*In" models introduced in the main text. The scenarios under study are: a static crowd in a waiting room; an intersection with flows crossing at 90° at low density; a bidirectional flow.

\end{document}